\documentclass{aastex}
\usepackage{emulateapj5}

\shorttitle{Long-Term X-Ray and TeV variability of Mrk 501}
\shortauthors{Gliozzi et al.}

  \def\xmm{{\it XMM-Newton}} 
  \def\asca{{\it ASCA}} 
   
  \def\rxte{{\it RXTE}} 
  \def\sax{{\it BeppoSAX}} 
   
  \def\rosat{{\it ROSAT}}

  \def\flux{erg cm$^{-2}$ s$^{-1}$}

  \def\ltsima{$\; \buildrel < \over \sim \;$}
  \def\simlt{\lower.5ex\hbox{\ltsima}} 
  \def\gtsima{$\; \buildrel > \over \sim \;$}
  \def\simgt{\lower.5ex\hbox{\gtsima}} 

\begin{document}
\title{Long-Term X-Ray and TeV variability of Mrk 501}

\author{M. Gliozzi}
\affil{George Mason University, 4400 University Drive, Fairfax, VA 22030}

\author{R.M. Sambruna}
\affil{NASA's Goddard Space Flight Center, Code 661, Greenbelt, MD 20771}

\author{I. Jung, H. Krawczynski}
\affil{Department of Physics, Washington University, St. Louis, MO 63130}

\author{D. Horan}
\affil{Fred Lawrence Whipple Observatory, Harvard-Smithsonian CfA, 
P.O. Box 97, Amado, AZ 85645-0097}

\author{F. Tavecchio}
\affil{INAF/Osservatorio Astronomico di Brera, via Bianchi 46, Merate and via 
Brera 28, Milano, Italy}

\begin{abstract}
We present X-ray observations of the nearby TeV blazar Mrk~501
obtained with RXTE during 1997, 1998, 1999, 2000, and 2004.  The goal
of this study is twofold: 1) characterize the long-term X-ray flux and spectral
variability of the source with a model-independent analysis, 
and 2) investigate the X-ray and TeV correlation on long timescales ($>$ days).
Significant spectral variations
were observed during all the observations along with long-term
timescale correlations between the X-ray colors and the count
rate.  Specifically,
on long timescales, a typical blazar behavior is observed with
the spectrum hardening when the source brightens, and the fractional
variability correlating with the energy band. A similar spectral trend
is observed also in the majority of the individual flares.
The spectral and temporal variability properties appear to be markedly different
compared to those of non-jet-dominated radio-loud and radio-quiet AGN monitored
with RXTE. To investigate the  X-ray -- TeV correlation on long timescales
we compared RXTE monitoring data with HEGRA and Whipple historical light curves. 
We confirm the presence of a direct
correlation between X-ray and $\gamma$-ray emissions, which appears to
be stronger when the source is brighter. The analysis of individual
flares shows that the X-ray -- TeV correlated activity is heterogeneous, in the
sense that it might be both linear and nonlinear, and some X-ray flares
seem to be lacking the TeV counterpart.
However, more sensitive TeV observations
are necessary to confirm these findings, and to put tighter constraints on
jet models. 
\end{abstract}


\section{Introduction}

BL Lacertae objects (BL Lacs) are members of the blazar class, which
are radio-loud AGN dominated by non-thermal continuum emission from
radio up to $\gamma$-rays (MeV to TeV energies) from relativistic
jets oriented at small angles to the observer 
(e.g., Urry \& Padovani 1995). 
The Spectral Energy Distributions (SEDs) of BL Lacs are
characterized by two components: one from radio to X-rays due to
synchrotron, generally peaking at UV/soft X-rays; and one peaking at
TeV gamma-rays, usually interpreted as inverse Compton scattering of ambient soft 
photons (e.g., Ulrich, Maraschi, \& Urry 1997).

Among BL Lacs, the nearby (z=0.034) source Mrk 501 is one of the
brightest at X-rays and a strong TeV emitter. This source came into
much attention after it exhibited a prolonged period of intense TeV
activity in 1997, accompanied by correlated X-ray emission on
timescales of days \citep{cata97,haya98,quin99,ahar97,djan99}.
Interestingly, this exceptional TeV activity was
accompanied by unusually hard X-ray emission up to $\gtrsim$ 100 keV
\citep{pian98,cata97,lame98,krawc00}, unprecedented in this or any other BL
Lac. The hard X-ray spectrum (photon index $\Gamma < 2$) implied a
shift toward higher energies of the synchrotron peak by more than
three decades, persistent over a timescale of $\sim$ 10 days (Pian et
al. 1998). In June 1998 a second TeV outburst was observed, again
accompanied by a shift of the synchrotron peak at higher energies on
timescales of 1 day or more \citep{samb00}. The peak shift is
positively correlated with the long-term luminosity changes \citep{tave01}. 

In this paper, we start a systematic study of the long-term X-ray flux 
and spectral
variability of Mrk~501 using proprietary and archival Rossi X-ray
Timing Explorer (\rxte) observations. 

The first part of this work 
is a model-independent study of the X-ray temporal and 
spectral properties of this jet-dominated source on long timescales. 
The main purpose of this analysis is to shed light on the role
played by the jet in the X-ray emissions of different classes of
AGN. Specifically, a model-independent characterization of the
temporal and spectral behavior of blazars (which are the only  AGN
with broad-band emission completely dominated by the jet), combined
with analogous information obtained from studies of radio-quiet AGN,
will allows one to investigate quantitatively the jet contribution in
radio-loud non-jet-dominated sources (e.g., Gliozzi, Sambruna \&
Eracleous 2003).  Once the jet contribution is properly assessed, the
physical parameters characterizing the accretion process onto the
supermassive black hole can be better constrained, and hence it is
possible to discriminate between competing theoretical models. This
model-independent characterization of the jet contribution at X-rays
is important not only for radio-loud AGN, but also for their
scaled-down counterparts, the Galactic black hole (GBHs), which  are
known to have compact radio jets during their ``low-hard"
spectral state. Indeed, despite the much higher signal-to-noise of
their spectra and light curves, the jet contribution to the total
X-ray emission in the low-hard state of GBHs is still matter of 
strong debate (see, e.g., Markoff et al. 2003;
Zdziarski et al. 2004).

Another goal of this work is to investigate the correlation between
the X-ray and TeV bands on timescales $>$ 1 day. TeV blazars, such as
Mrk~501, are known to show a positive correlation between X-ray and
TeV emission, although exceptions exist (e.g., Krawczynski et
al. 2004). Detailed studies of the X-ray -- TeV correlation are
crucial to constrain and discriminate the two competing classes of
models: the leptonic models, where the TeV emission is due to the
synchrotron self-Compton (SSC) process (e.g. Tavecchio et al. 2001),
and the hadronic models, in which the TeV emission is the
product of cascades initiated by relativistic hadrons.
More specifically, in hadronic models the continuum emission is explained 
by hadronic interactions of a highly relativistic baryonic outflow which 
sweeps up ambient matter \citep{Pohl:00}, by interactions of high-energy 
protons with gas clouds moving across the jet \citep{Dar:97}, or by interactions 
of ultra high energy  protons with ambient photons \citep{Mann:93}, 
with the jet magnetic field \citep{Ahar:00a}, or with both \citep{Muec:02}.

The outline of the paper is as follows. In $\S~2$ we describe the
observations and data reduction. In $\S~3$ we study the X-ray
variability of Mrk~501 during long-term monitorings and individual flares,
using the fractional variability to quantify the temporal behavior  
and the hardness ratios to investigate the spectral variability.  In
$\S~4$ we investigate the X-ray -- TeV correlation using
data from HEGRA and Whipple in combination with RXTE observations. 
In $\S~5$ we discuss the implications of the results of the
temporal and spectral analysis. Finally,
we summarize the main conclusions in $\S~6$.  For completeness, in
Appendix 1 we briefly report the main results from a time-averaged
spectral analysis, and in Appendix 2 we summarize the results from the
temporal and spectral analyses of July 1997.
\section{Observations}

\subsection{Sampling}

\noindent\emph{1997:} \rxte\ observed Mrk~501 in the 2--60 keV energy
band (as in all others observations with the RXTE PCA) beginning on
March 18 with the following pattern: the source was observed on March
18 and then once or twice a day from April 3--16, May 2--15, and July
12--16.  During this period there was a large outburst of $\gamma$
rays extending to energies greater than 20 TeV (Aharonian et
al. 1997,1999).\\

\noindent\emph{1998:} \rxte\ observed Mrk~501 as part of different
campaigns by different PIs starting February 25 with the following
sampling: throughout the monitoring campaign the source was observed
regularly every three or four days with the exception of three periods
(May 15--29, June 15--29, and July 23--29) when the sampling density
was increased (one to five observations per day).  
The TeV data of Mrk~501 were obtained by the Cherenkov telescope
system of the High-Energy Gamma-Ray Observatory (HEGRA).
Unfortunately, the source was in a low state at TeV gamma-rays, except
for two weeks in June 1998, when it underwent a large burst at both
X-rays and TeV energies \citep{samb00,ahar01}.\\

\noindent\emph{1999:} \rxte\ observed Mrk~501 with the following
pattern: there were three periods of dense sampling (one to three
observations per day) on the dates May 6--26, June 4--10, and July
4--5.  The source was weak at TeV energies with an average integral
flux 13 times lower than that of 1997 \citep{ahar01}.\\

\noindent\emph{2000:}
Starting July 20, 2000 \rxte\ observed Mrk 501 with the
following pattern: for three times the source was monitored
daily for 14 days, with a gap of two weeks in between the observation
periods. 
A portion of the total allocated \rxte\ observing time was
used earlier in the year to monitor Mrk 421 in flaring state
\citep{krawc01}. 
During the 2000 X-ray monitoring Mrk~501 was very weak at TeV, with no
episodes of major outbursts.\\  

\noindent\emph{2004:}
In addition to the long-term monitoring campaigns, whose data have been
at least partially published, we report on a recent X-ray -- TeV flare
observed in June 2004.
Starting from June 11 2004, Mrk~501 was observed with \rxte\ for thirteen
days and with Whipple for eight days. Unfortunately the source was caught during a
period of dim TeV activity, yielding mostly upper limits in that
energy range and hence hampering the investigation of the 
X-ray -- $\gamma$-ray correlation.\\

\begin{figure}
\begin{center}
\includegraphics[bb=45 60 550 730,clip=,angle=90,width=16cm]{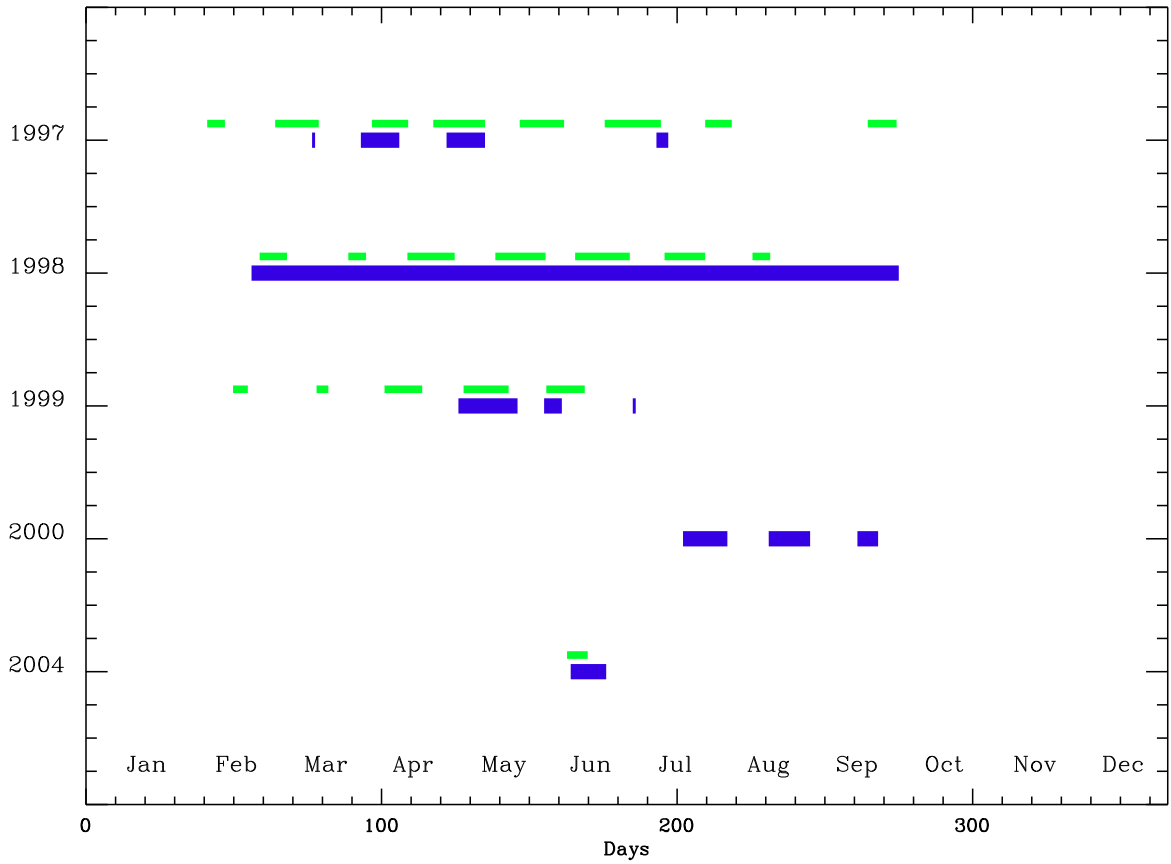}

\end{center}
\caption{Summary of the monitoring campaigns of Mrk~501 in the X-ray 
(darker thicker line) and in the TeV (lighter thinner line) energy ranges. 
The continuous intervals do not mean that the observations are actually uninterrupted, 
but simply that they have been carried out with a sampling typical for each instrument, which is
3--4 days for \rxte\ observations and 1--2 days for the TeV observations
performed with HEGRA/Wipple. Specifically, in this plot any temporal gap 
shorter than 5 days has been neglected (i.e., considered as continuous) in the \rxte\
monitoring. Similarly, any gap shorter than 3 days has been neglected in the TeV
monitoring.}
\label{figure:fig1}
\end{figure}

\begin{figure}
\includegraphics[bb=30 165 561 655,clip=,angle=90,width=16.cm]{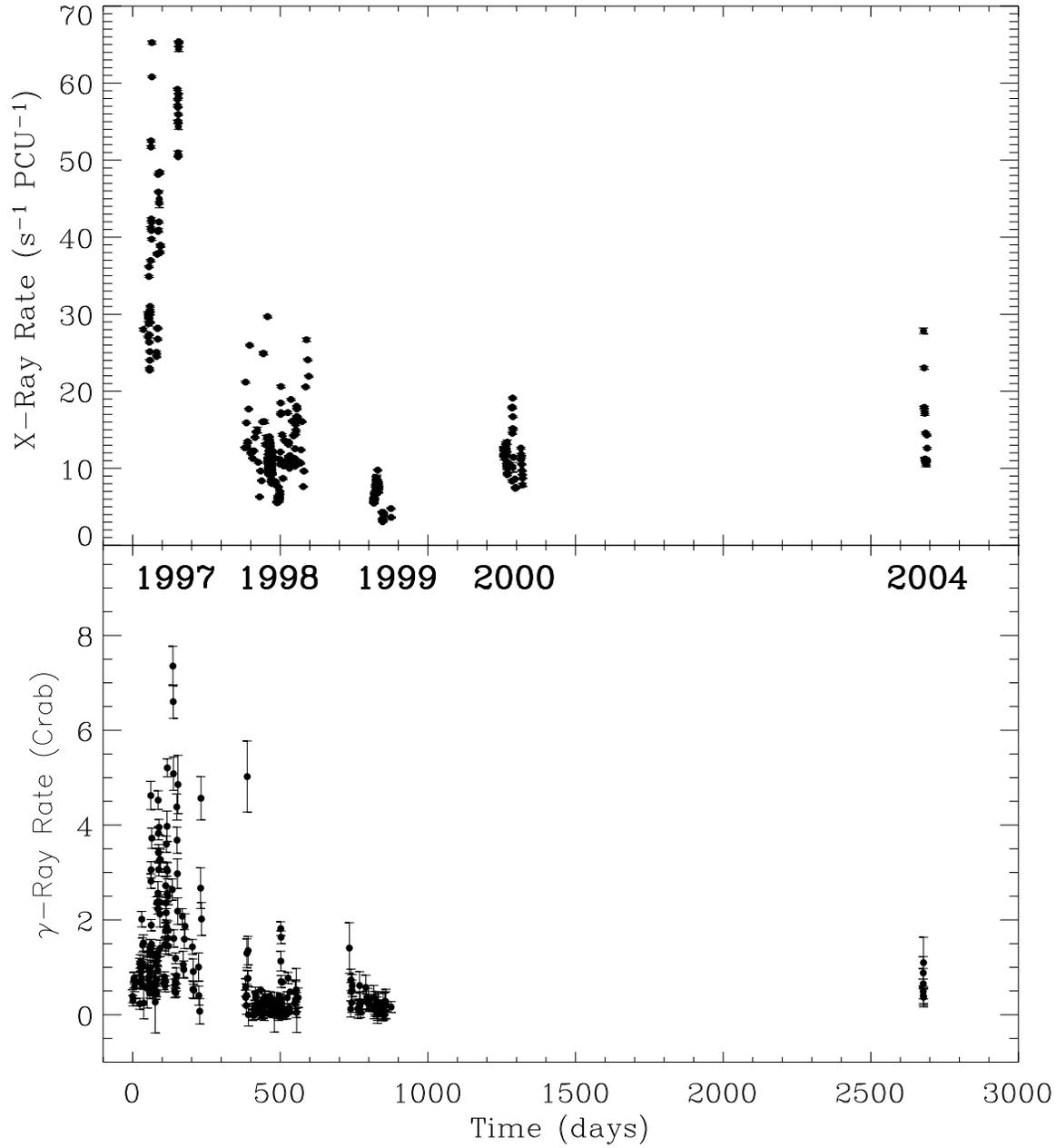}
\caption{Historical X-ray (top panel) and TeV (bottom panel)
light curves of Mrk~501 starting on February 2nd 1997 (MJD=50489.5)
and finishing on June 24th 2004 (MJD=53180.3).  X-ray time bins are 5760 s ($\sim$ 1 \rxte\ orbit).} 
\label{figure:fig2}
\end{figure}

Figures~\ref{figure:fig1} and~\ref{figure:fig2} describe the monitoring 
campaigns of Mrk~501. In Figure~\ref{figure:fig1} we summarize the temporal 
coverage in the X-ray (darker thicker line) and in the TeV (lighter thinner 
line) energy bands, carried out by \rxte\ and HEGRA/Whipple, respectively. 
Continuous intervals do not mean that the observations are actually 
continuous, but simply that they have been carried out with samplings typical 
for each instrument,
which means every 3--4 days for \rxte\ observations and 1--2 days for the TeV 
observations performed with HEGRA/Whipple. As a consequence,
in this figure any temporal gap 
shorter than 5 days has been neglected (i.e., considered as continuous) in 
the \rxte\ monitoring. Similarly, any gap shorter than 3 days has been 
neglected in the TeV monitoring.

Figure~\ref{figure:fig2} shows the historical X-ray and TeV light curves 
starting February 2nd 1997 (MJD=50489.5) and ending  June 24th 2004 
(MJD=53180.3). The light curves reveal the significant change of count rate, 
which occurred in both energy bands over the seven-year monitored interval.
As an operative criterion for this work, we consider ``contemporaneous" 
X-ray -- TeV observations those taken less than 8 hours apart. The source
may vary on much shorter timescales, which could complicate some of the
interpretations.

\subsection{Data reduction} 

\noindent{\bf X-rays:} The PCA and HEXTE data were reduced using
\verb+REX+, with standard screening criteria for a \verb+FAINT+
source. Some of the proportional counter units (PCUs) were
occasionally turned off; we used all the available PCUs and then
normalized the count rate to 1 PCU.  The current temporal analysis is
restricted to Proportional Counter Array (PCA; Jahoda et al. 1996),
STANDARD-2 mode, 2--20 keV, Layer 1 data, because that is where the
PCA is best calibrated and most sensitive. The data were initially
extracted with 16 s time resolution and subsequently re-binned at
different bin width depending on the purpose for which they were
employed. In most of the cases we used a binning time of the order of the
orbital timescale of \rxte\ (5760 s).
The time-averaged spectral analysis was performed using PCA data in the
4--20 keV band and HEXTE data in the 20--60 keV range, where the instruments
are better calibrated and the signal-to-noise ratio is higher. 
Spectral analysis of the data was performed using \verb+XSPEC+
v. 11.3.1. \\
\vskip 0.1cm

\noindent{\bf TeV  Observations:}\\
We used a large data base of published TeV gamma-ray data
(e.g., Quinn et al. 1999; Aharonian et al. 2001). 
In addition, new data were acquired with the Whipple
10 m Cherenkov telescope (Mount Hopkins, AZ). The Whipple collaboration
kindly allowed us to use the data for this publication. 
Detailed descriptions of Whipple
observing modes and analysis procedures can be found elsewhere
(e.g., Punch et al. 1992; Reynolds et al. 1993; Weekes 1996). 
Based on the Whipple measurements of the energy spectrum from the Crab
nebula \citep{hill98}, a flux of 1 Crab corresponds to a differential 1
TeV flux of (3.20$\pm$0.17$_{\rm stat}$$\pm$$0.6_{\rm syst}$)
$\times10^{-11}$ photons cm$^{-2}$ s$^{-1}$ TeV$^{-1}$ and to a $\nu
F_\nu$ flux of (5.12$\pm$0.27$_{\rm stat}$$\pm$$0.96_{\rm syst}$)
$\times10^{-11}$ ergs cm$^{-2}$ s$^{-1}$.  The upper limits were computed
with the Bayesian formalism described by Helene et al. (1983).

\section{X-ray Temporal Analysis}
Here we investigate the temporal and spectral variability properties
of Mrk~501 during  long-term monitorings (from weeks to several months) 
and individual flares  using different  model-independent techniques, 
based on X-ray colors and fractional variability.  

\begin{figure}
\centering
\includegraphics[bb=70 20 400 375,clip=,angle=0,width=7cm]{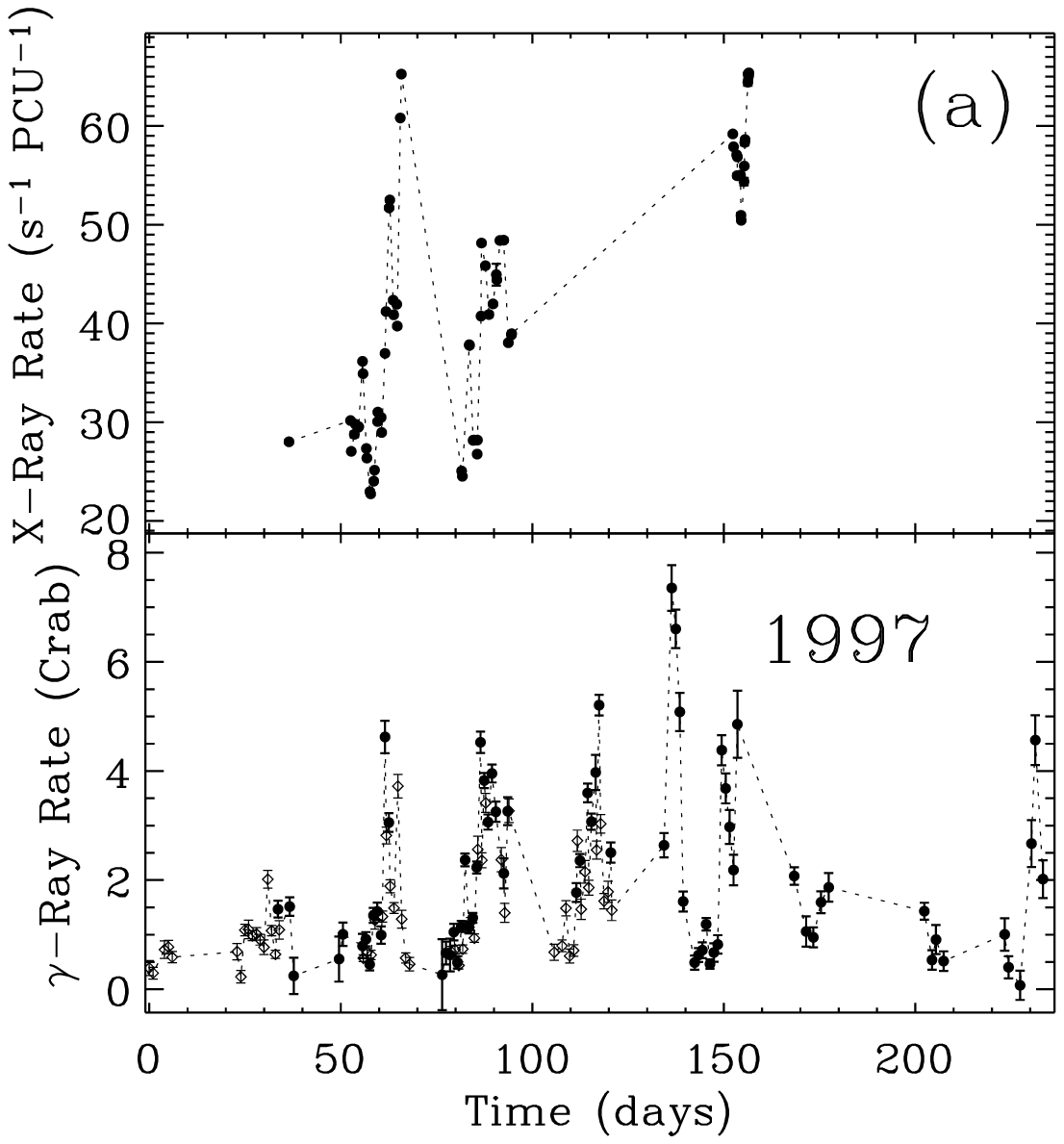}\includegraphics[bb=70 20 400 375,clip=,angle=0,width=7.cm]{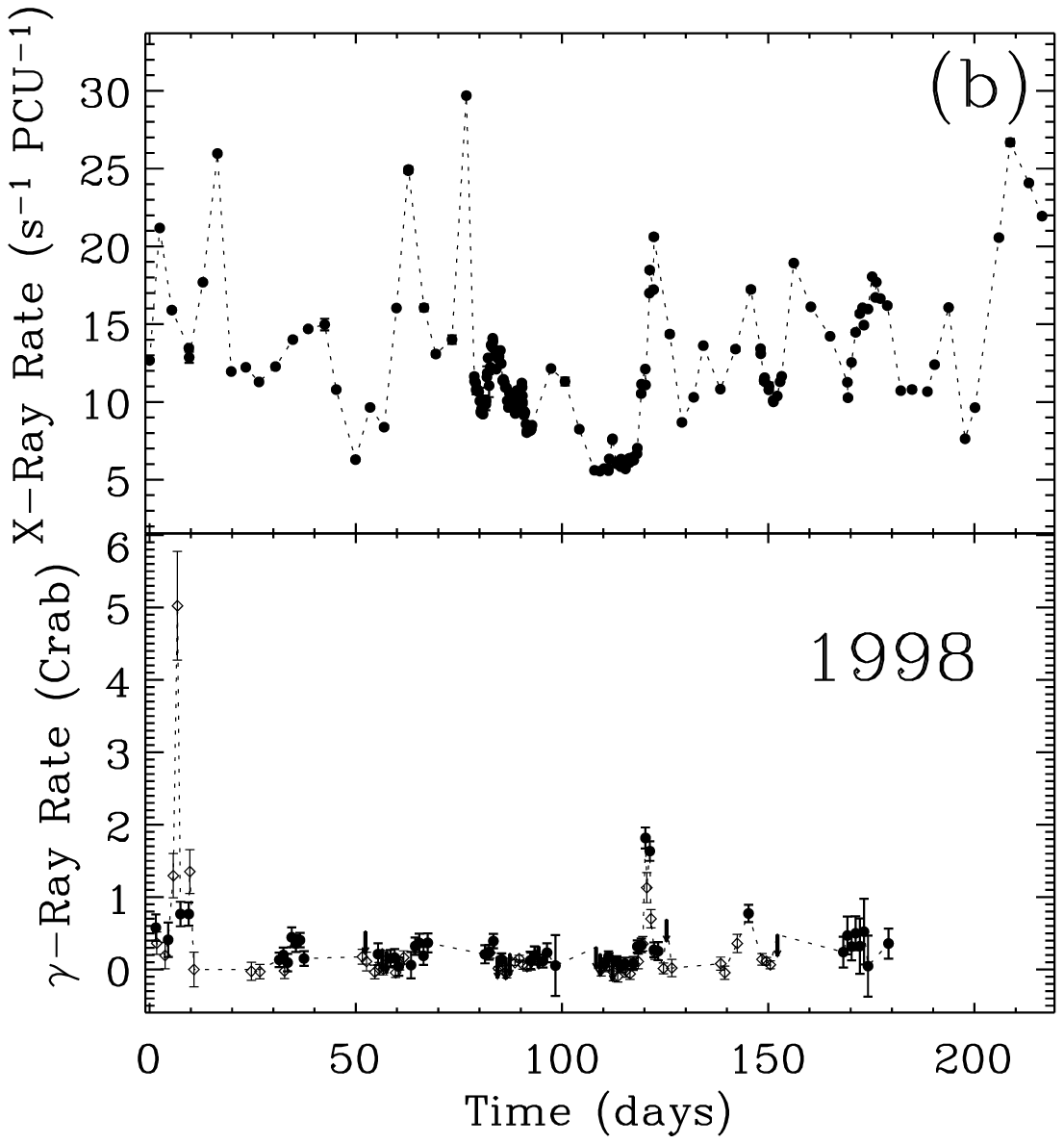}
\includegraphics[bb=70 20 400 375,clip=,angle=0,width=7.cm]{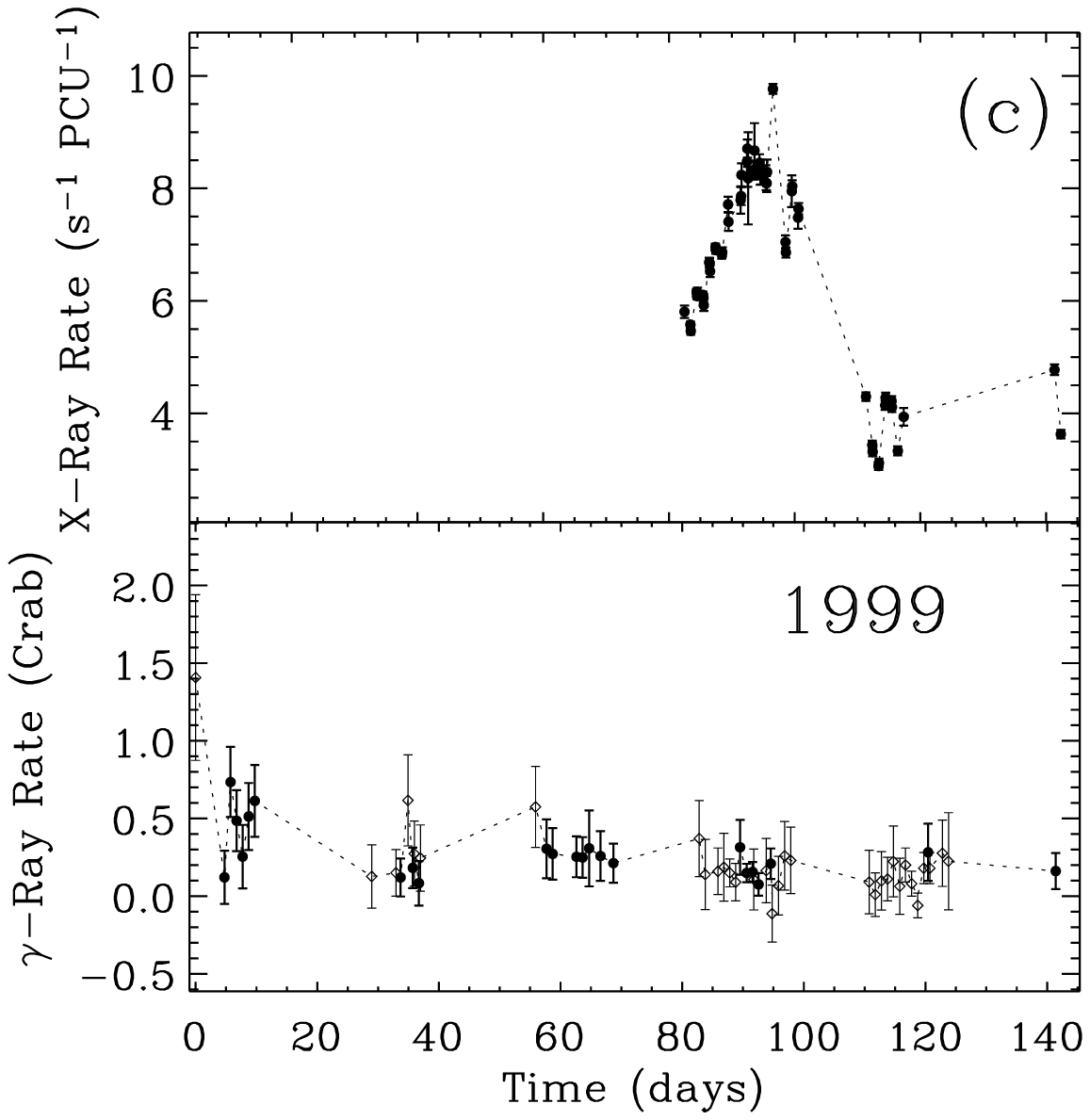}\includegraphics[bb=70 20 400 375,clip=,angle=0,width=7.cm]{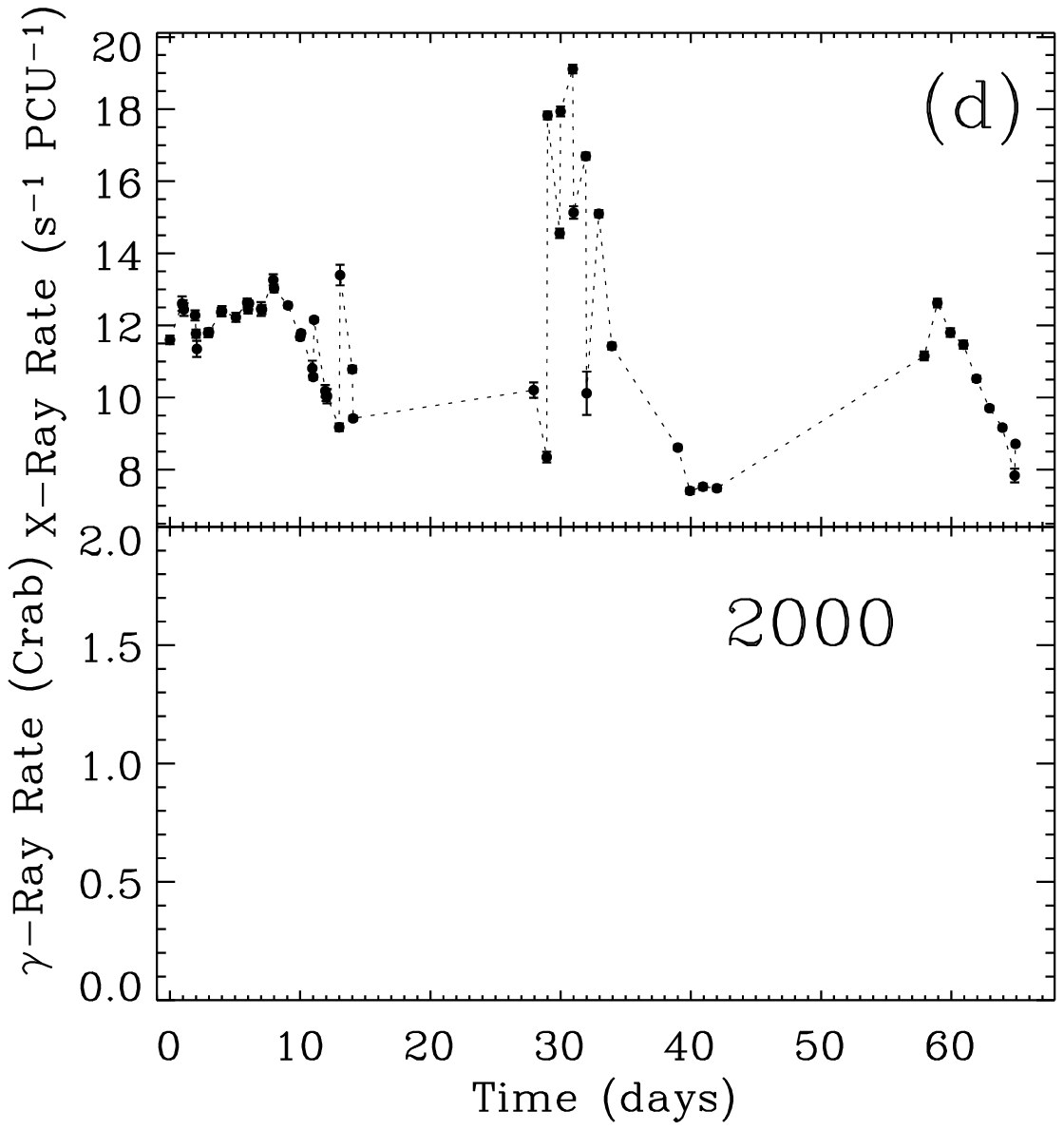}
\includegraphics[bb=70 20 400 375,clip=,angle=0,width=7.cm]{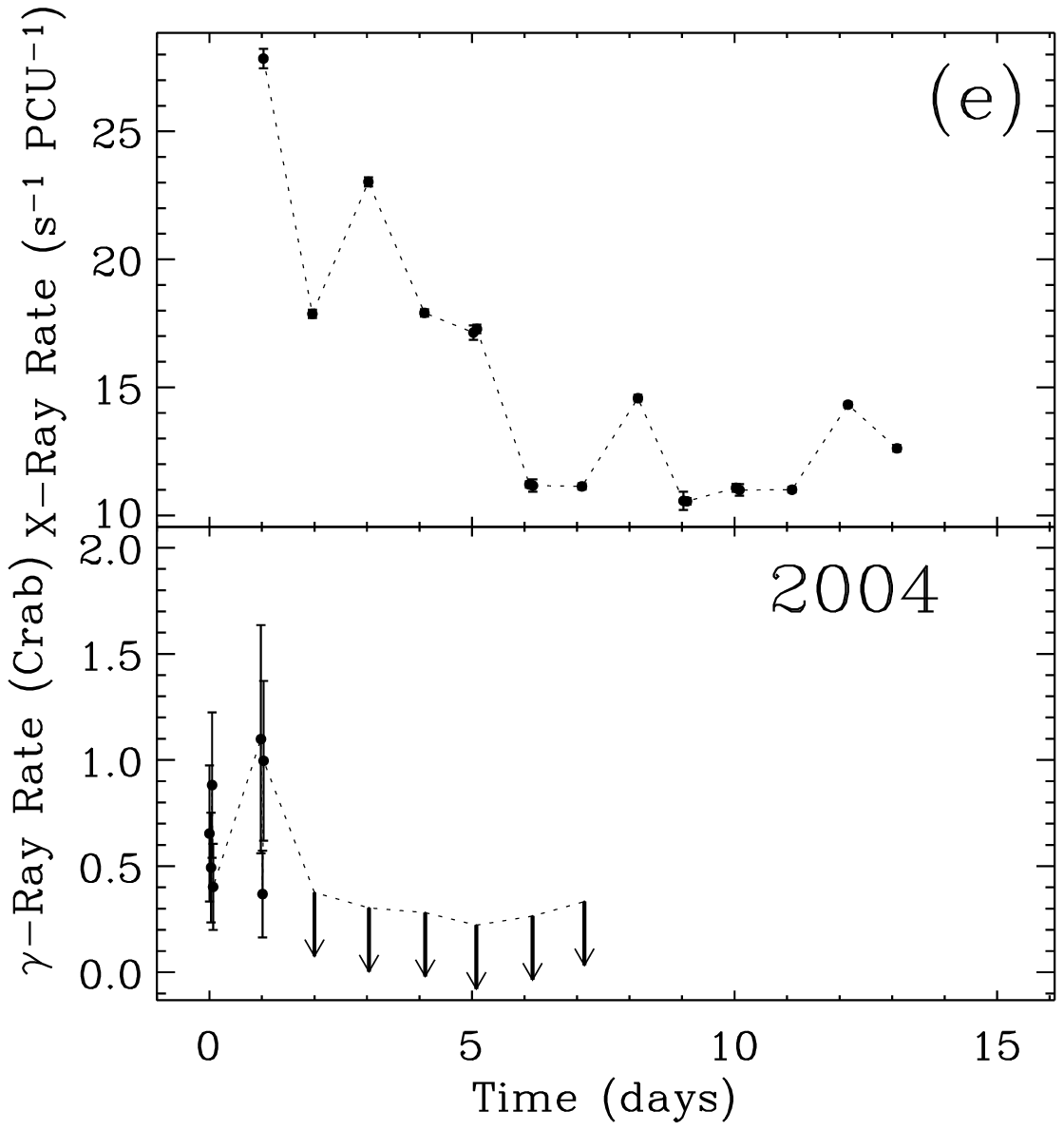}
\caption{$\gamma$-Ray and 2-20 keV X-ray light curves
during the 1997 (a), 1998 (b), 1999 (c), 2000 (d),
and 2004 (middle right) multiwavelength campaigns, respectively. In the $\gamma$-ray light curves  
filled dots correspond to HEGRA measurements, open diamonds to Whipple observations, 
and the arrows to upper limits. X-ray time bins are 5760 s ($\sim$ 1 \rxte\ orbit).
}
\label{figure:fig3}
\end{figure}

Figure~\ref{figure:fig3} 
shows the background-subtracted light curves
of Mrk~501 in the 2--20 keV energy band, along with the
coordinated TeV observations.
The X-ray count rate is normalized to 1 PCU and
the time binning is one \rxte\ orbit ($\sim$ 5760 s).
The first relevant feature is the significant difference in the mean
count rate (note the different scale on the y axes of Fig.~\ref{figure:fig3}) 
from one year to another. 
This can be better appreciated comparing the total count rates and using a 
uniform y range, as done in the top panel of Figure~\ref{figure:fig2}.
The 2--20 keV mean count rate decreases from 41.5 c/s/PCU in 1997, to 
11.6 c/s/PCU in 1998, reaching a minimum of 6.6 in 1999, and then rises again
at 11.7 c/s/PCU in 2000, and 14.7 c/s/PCU in 2004.
All light curves are characterized by
frequent flares with variable amplitude. Typically, the count rate
in the 2--20 keV energy band changes by factors 2--3 on timescales of
5--10 days.

\subsection{X-Ray Colors}
Light curves in 2--5 keV (soft), 5--10 keV (medium), and 
10--20 keV (hard) energy bands were extracted to investigate whether the 
observed flux variability is accompanied by spectral variability. 
We found significant and usually coordinated variability between energy bands.
Starting from the light curves in the three selected energy band, we
defined two X-ray colors as ratios of count rates
in different bands: ${\rm Hard/Soft}=F_{10-20}/F_{2-5}$, and ${\rm
Med/Soft}=F_{5-10}/F_{2-5}$. X-ray colors provide a simple, model-independent 
way to investigate spectral variability.

We carried out $\chi^2$ tests against the hypothesis of constancy for the 
X-ray color light curves to assess quantitatively the presence of spectral 
variability during the different monitoring campaigns of Mrk~501.
For all the cases we found
a chance probability $\ll 0.1\%$, indicating the presence of a strong
spectral variability.  

\begin{figure}
\centering
\includegraphics[bb=65 143 278 330,clip=,angle=0,width=5.5cm]{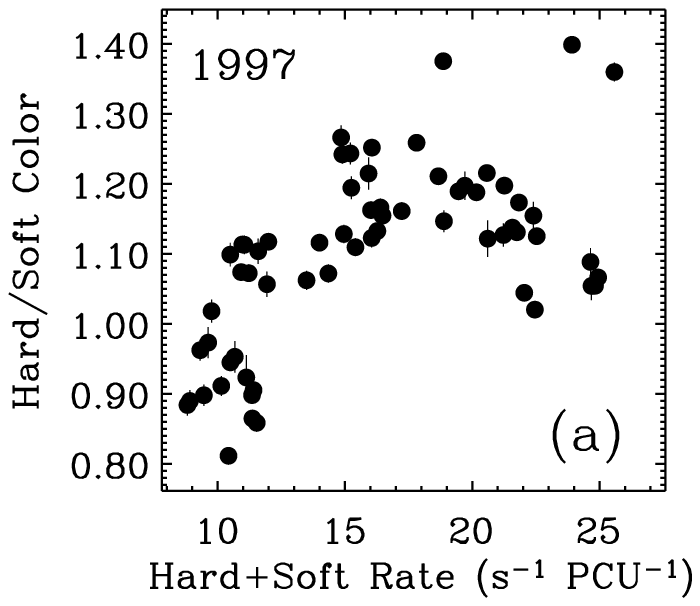}\includegraphics[bb=65 143 278 330,clip=,angle=0,width=5.5cm]{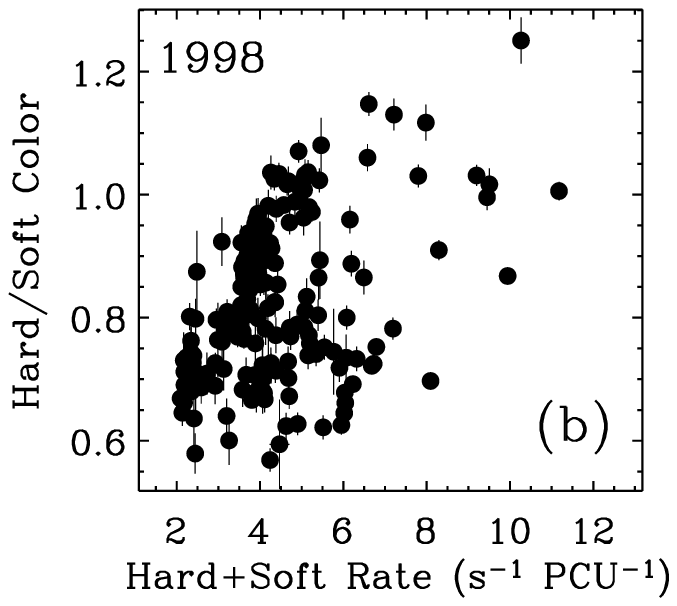}\includegraphics[bb=65 143 278 330,clip=,angle=0,width=5.5cm]{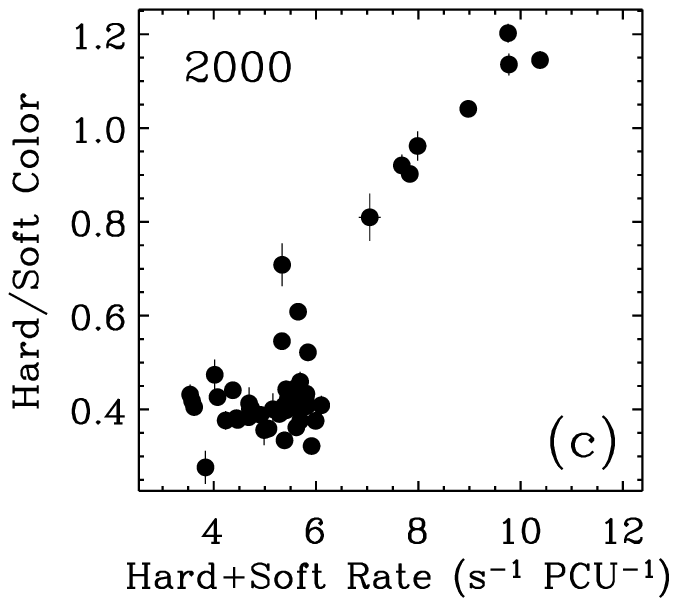}
\includegraphics[bb=65 143 278 330,clip=,angle=0,width=5.5cm]{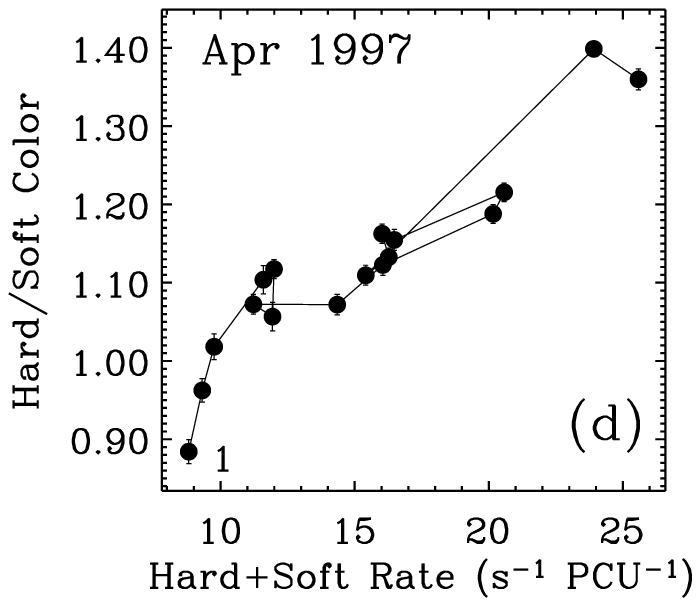}\includegraphics[bb=65 143 278 330,clip=,angle=0,width=5.5cm]{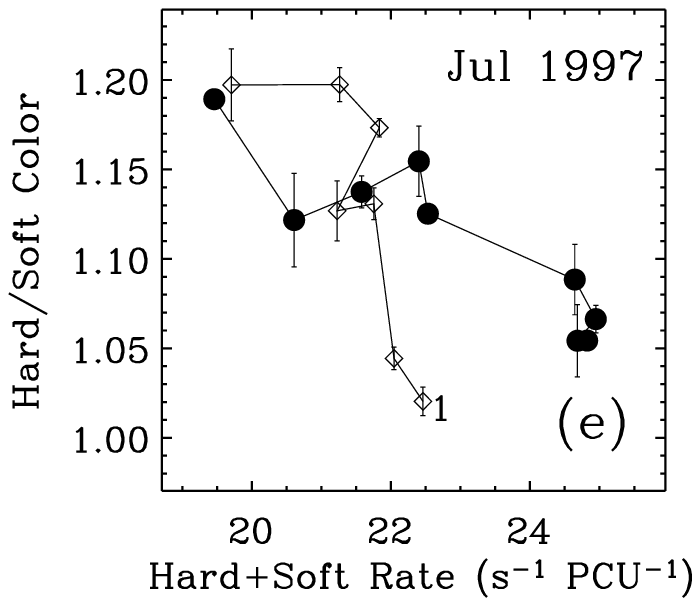}\includegraphics[bb=65 143 278 330,clip=,angle=0,width=5.5cm]{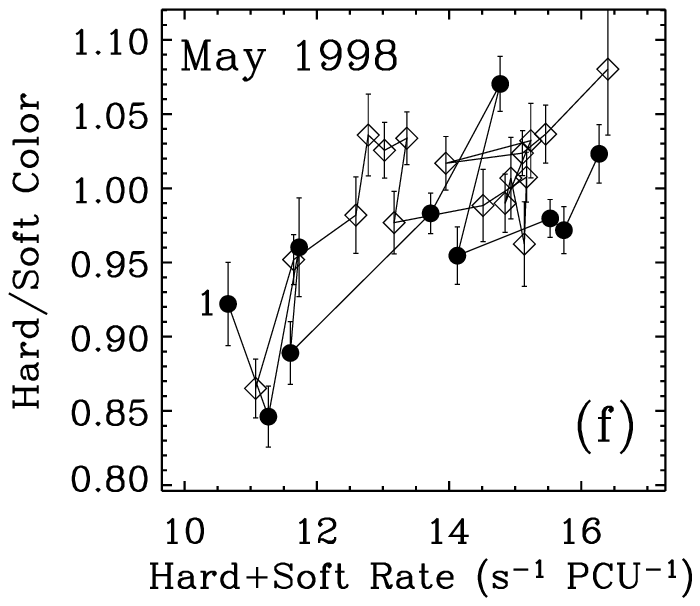}
\includegraphics[bb=65 143 278 330,clip=,angle=0,width=5.5cm]{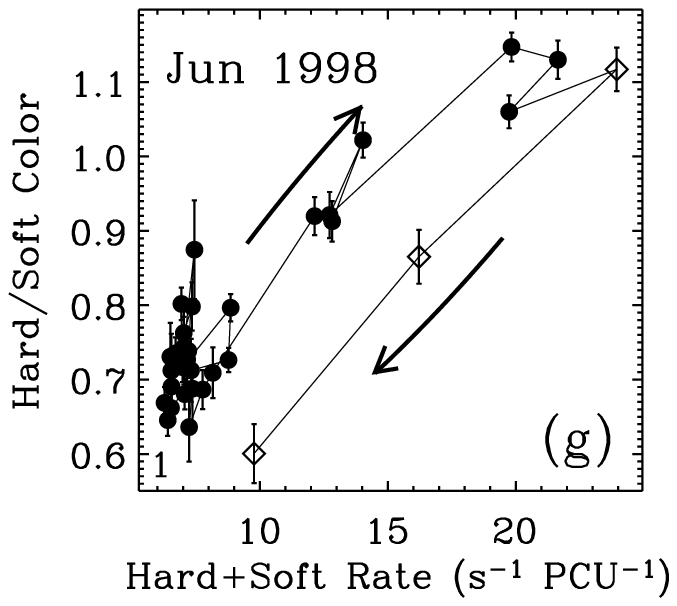}\includegraphics[bb=65 143 278 330,clip=,angle=0,width=5.5cm]{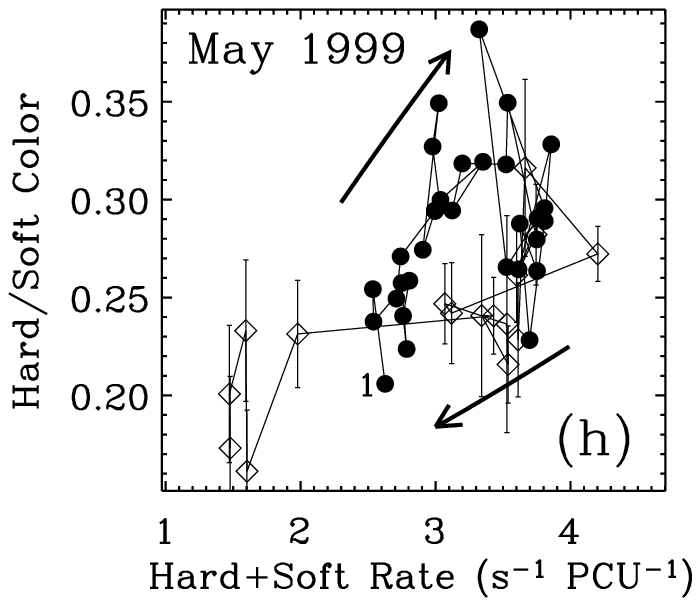}\includegraphics[bb=67 143 278 330,clip=,angle=0,width=5.5cm]{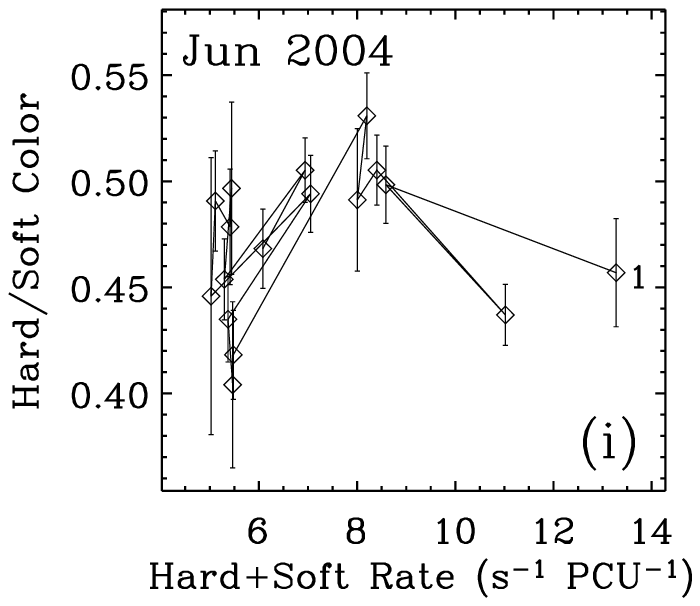}
\caption{X-ray color [10-20 keV]/[2-5 keV] plotted versus count rate for the total light curves 
in 1997 (a), 1998 (b), 2000 (c) 
and for individual flares (d-i). For individual flares
filled circles represent points during the flare rise and open diamonds
represent points during the decay phase. Time bins are 5760 s ($\sim$ 1 \rxte\ orbit).
For the sake of clarity in Fig.4h error-bars have been drawn only for the decay phase.}
\label{figure:fig4}
\end{figure}

Figure~\ref{figure:fig4} shows the Hard/Soft X-ray color plotted
versus the count rate. Panels a, b, and c 
show the three long monitoring campaigns carried out
during 1997, 1998, and 2000; we do not consider the observations
in 1999 and 2004 as long-term monitorings, since they
cover only one flare. Figures~\ref{figure:fig4} d-i
show the hardness ratio - count rate (hereafter, $HR-ct$) plots for
several prominent flares, with the rising phase described by filled circles
and the decay phase by open diamonds. The number $1$ represents the starting
point. We restrict our analysis to flares
with the best coverage, meaning those with at least 10--15 data points
in the X-ray light curve (assuming time-bins of one satellite's orbit)
and, possibly, with associated TeV emission.

A visual inspection of Figures~\ref{figure:fig4} a, b, and c suggests 
the presence of a positive
trend, within a considerable scattering, with the source hardening when
the count rate increases. To verify whether this apparent trend is 
statistically significant and quantify the degree of
correlation, we performed the following tests. First, we computed 
the Pearson's 
linear-correlation coefficient {\it r} and the relative chance 
probability $P_{\rm Pearson}$. Since in some cases (for example if the 
underlying 
correlation is not linear) the Pearson's correlation coefficient might be
misleading, we also computed the degree of correlation using two non-parametric
statistics, such as the Spearman's rank $\rho$ and the  Kendall's rank $\tau$.
In addition, we estimated the best-fit linear slope using a least square
fitting method. 
The results of the correlation analysis are reported in Table~1 and confirm,
at high significance level, that $HR$ is positively correlated with the count
rate during the three long-term monitoring campaigns. Similar conclusions
are derived also when the X-ray color  ${\rm Med/Soft}=F_{5-10}/F_{2-5}$ is
used instead of ${\rm Hard/Soft}=F_{10-20}/F_{2-5}$.

The same analysis carried out on individual flares leads to similar
conclusions: the flares harden during the rising phase and
soften during the decay phase. A noticeable exception is
the event on July 1997, where the opposite trend is observed
(Fig.~\ref{figure:fig4}e). 
This spectral behavior, 
very unusual for a blazar and hard to reconcile with the simplest
blazar models, was first noticed by Lamer \& Wagner (1998), who analyzed
the \rxte\ observations of Mrk~501 in July 1997.
Note that
this event is not a single flare, but is composed of 
a decay phase of a first flare and the rising phase of second flare. 
We discuss this result in more detail in $\S 5$.

Frequently, in other well studied HBL (for recent studies based on 
high-quality \xmm\  data of Mrk 421 see Brinkmann et al. 2003; 
Ravasio et al. 2004)
the inspection of individual flares in the hardness ratio
- count rate plot exhibit hysteresis loops, both in a clockwise and
anti-clockwise sense. Figures~\ref{figure:fig4}g and \ref{figure:fig4}h 
(for which arrows have been drawn to guide the eye)
suggest the presence
of clockwise hysteresis loops during June 1998 and May 1999,
where the points
corresponding to the rising phase of the flares (filled circles) are located
at higher values of $HR$ compared to points of the decay phase (open
diamonds) of similar count rate level. In July 1997 (Fig.~\ref{figure:fig4}e), 
the presence of an anti-clockwise loop is an artifact, in the sense that it is 
due to the combination of two different flares. In all other cases 
(Figures~\ref{figure:fig4}d,f,i) no signs of hysteresis loops are present.

In summary, all results consistently indicate the presence of a positive 
correlation between $HR$ and $ct$ (within a considerable scattering) 
during the long-term monitoring campaigns
and for nearly all the flares investigated (the only clear exception is July 1997
and will be discussed in $\S$5).
This is a typical behavior for 
high-energy peaked BL Lacs (HBL): the spectrum hardens when the source gets brighter.

\subsection{Fractional Variability}

A simple way to quantify the variability properties of Mrk~501,
without considering the time ordering of the values in the
light curves, is based on the fractional variability parameter
$F_{\rm var}$ (e.g. Rodriguez-Pascual et al. 1997; Edelson et
al. 2002; Vaughan et al. 2003). This is a common measure of the intrinsic variability
amplitude relative to the mean count rate, corrected for the effect of
random errors, i.e.,
\begin{equation}
F_{\rm var}={(\sigma^2-\Delta^2)^{1/2}\over\langle r\rangle}
\end{equation}
\noindent where $\sigma^2$ is the variance, $\langle r\rangle$ the
unweighted mean count rate, and $\Delta^2$ the mean square value of
the uncertainties associated with each individual count rate.
The square of the fractional variability parameter, $F_{\rm var}^2$,
is called normalized excess variance
$\sigma_{\rm NXS}^2$, whereas the excess variance
$\sigma_{\rm XS}^2$ is defined as 
 $\sigma_{\rm NXS}^2\times\langle r\rangle^2$.

\begin{figure}
\centering
\includegraphics[bb=77 149 265 320,clip=,angle=0,width=5.5cm]{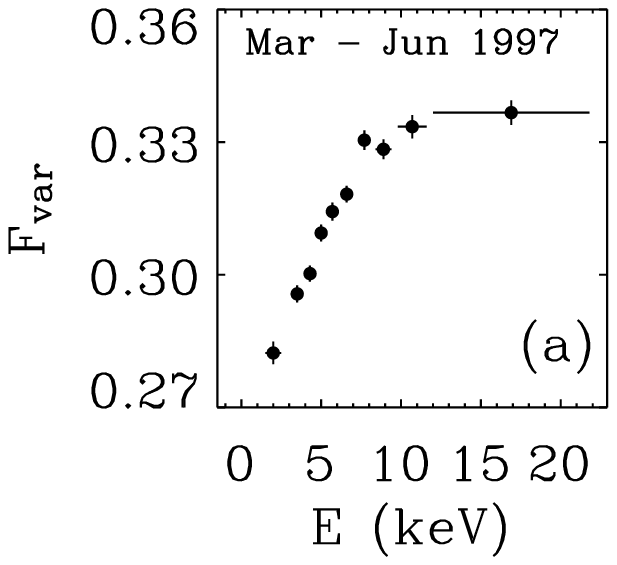}\includegraphics[bb=77 149 265 320,clip=,angle=0,width=5.5cm]{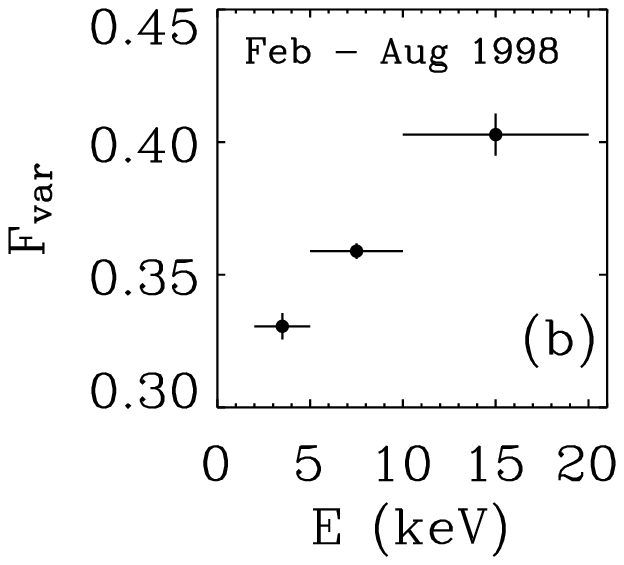}\includegraphics[bb=77 149 265 320,clip=,angle=0,width=5.5cm]{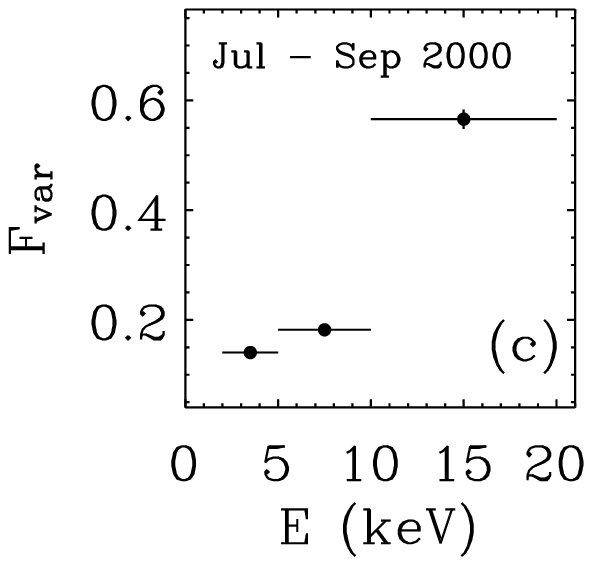}
\includegraphics[bb=77 149 265 320,clip=,angle=0,width=5.5cm]{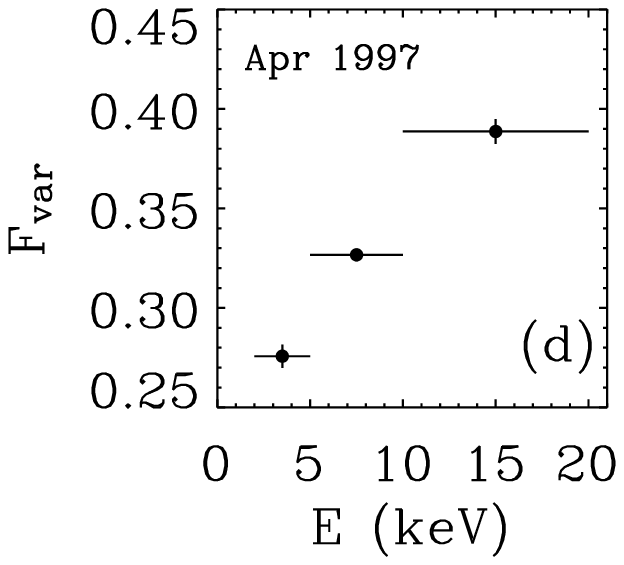}\includegraphics[bb=77 149 265 320,clip=,angle=0,width=5.5cm]{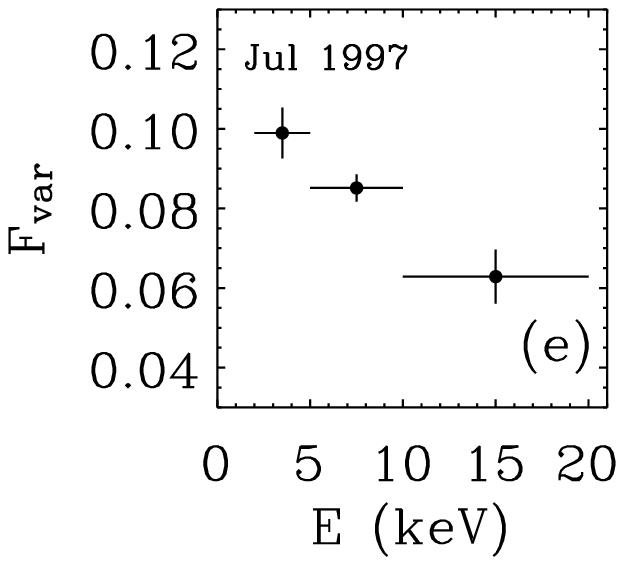}\includegraphics[bb=77 149 265 320,clip=,angle=0,width=5.5cm]{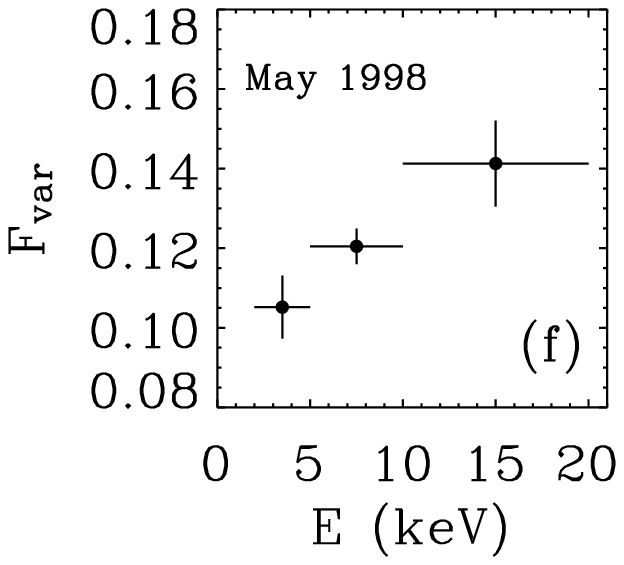}
\includegraphics[bb=77 149 265 320,clip=,angle=0,width=5.5cm]{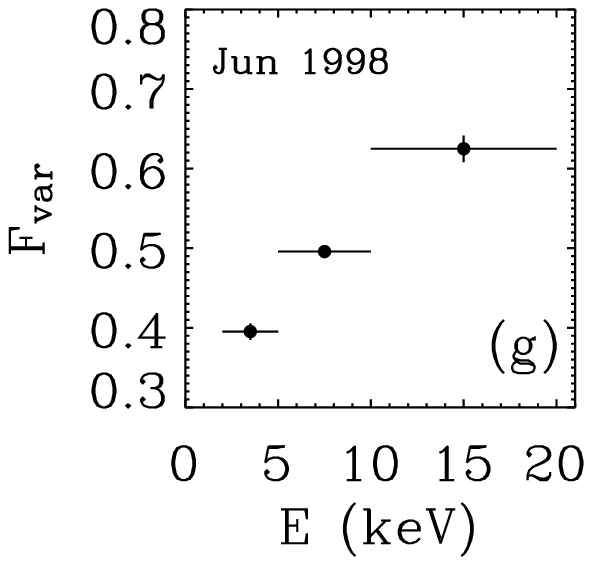}\includegraphics[bb=77 149 265 320,clip=,angle=0,width=5.5cm]{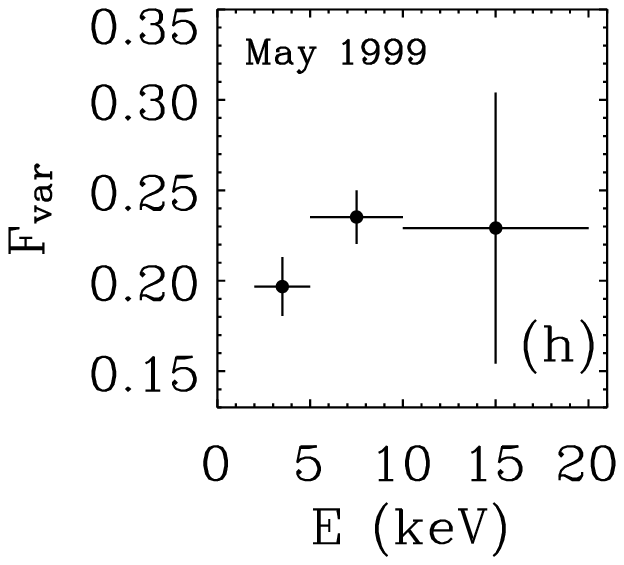}\includegraphics[bb=77 149 265 320,clip=,angle=0,width=5.5cm]{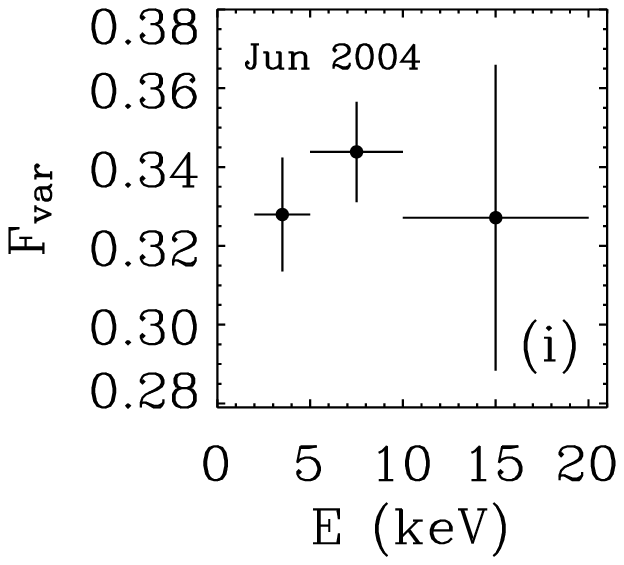}
\caption{Fractional variability parameter plotted versus the
energy for the total light curves in 1997 (a), 1998 (b), 2000 (c) 
and for individual flares (d-i). The error-bars along the x axis simply represent
the energy band width. The error-bars along the y axis represent the 3$\sigma$ value computed following
Vaughan et al. 2003.} 
\label{figure:fig5}
\end{figure}

We carried out two different kind of analyses using the fractional
variability parameter: 1) $F_{\rm var}$ vs. $E$ and 2) $F_{\rm var}$ vs. $ct$.
In the first kind of analysis, we computed $F_{\rm var}$ on selected
energy bands for the whole 1997, 1998, and 2000 light curves, and for
the individual flares selected above. We
used light curves in three energy bands, 2--5 keV (soft band), 5--10
keV (medium band) and 10--20 keV (hard band), and calculated $F_{\rm
var}$ for all of them.  All bands show significant variability. The
very high count rate of Mrk~501 in 1997 allows one to perform the
same analysis using a larger number (10 instead of 3) of narrower
energy bands. For
all long-term observations, we found suggestive evidence that the amplitude 
of variability increases with increasing energy band 
(see Fig.~\ref{figure:fig5} a,b,c). Similar positive trends are
found in most of the flares, again with the exception of July 1997
(Fig.~\ref{figure:fig5}e)
which shows an opposite trend, whereas flares in May 1999 
(Fig.~\ref{figure:fig5}h) and June 2004 (Fig.~\ref{figure:fig5}i)
are basically consistent with the hypothesis of constant variability
in all bands. 

The results of a quantitative correlation analysis (performed with methods
described in the previous section) are summarized in Table~2, and confirm
the presence of a positive trend with increasing fractional variability
with increasing energy during the long-term monitoring
campaigns and for three flares. Since, with the exception of 1997 campaign, 
the $F_{\rm var}\,-\,E$ plots account for three points only (each of which accounts 
for information from energy-selected light curves), the significance of the 
statistical tests might be questionable. However, it is worth noting that the 
highly significant positive correlation found in 1997 (with 10 data points) is not 
due to a particularly
steep trend in the $F_{\rm var}\,-\,E$ plane (in fact the 1997 slope is flatter 
than those in 1998 and 2000), but simply to the fact that the better photon
statistics allowed us to compute $F_{\rm var}$ over a larger number of energy 
bands. Therefore, it is reasonable to expect that also in 1998 and 2000 a
strong positive correlation would have been confirmed, if $F_{\rm var}$ could have
been computed on several energy bands.
In conclusion, the fact that a positive trend
is found in all long-term monitorings and in most of the flares,
along with the fact that a positive trend still exists when
$F_{\rm var,soft}$
is increased by a factor 3$\sigma$ and $F_{\rm var,hard}$ is decreased by
3$\sigma$, strongly argues in favor of a genuine positive correlation.

In the second kind of analysis, we
calculated $F_{\rm var}$ over selected time-intervals, in order to
look for the presence of a correlation (or anti-correlation) with the
mean count rate. Since the value of $F_{\rm var}$ depends both on the
duration of the interval and on the number of data points, we chose
intervals of similar length and with a time binning appropriate to
produce a similar number of points in each interval.  This analysis
was carried out both on long and short timescales: in the former
case, we selected several intervals over the entire period spanned by
the monitoring campaigns (i.e., from March 1997 to July 2000), whereas
to probe shorter timescales we focused on May 1998, when the monitoring
frequency was the highest. 
Specifically, to investigate the long-term $F_{\rm var}$ -- ct correlation, we
selected intervals of duration of $\sim$13--15 days with at least one
observation per day, and re-binned the light curves at 1 orbit.
With these criteria, we were able to select
seven intervals of similar length and sampling.

\begin{figure}
\centering
\includegraphics[bb=46 36 350 435,clip=,angle=0,width=8cm]{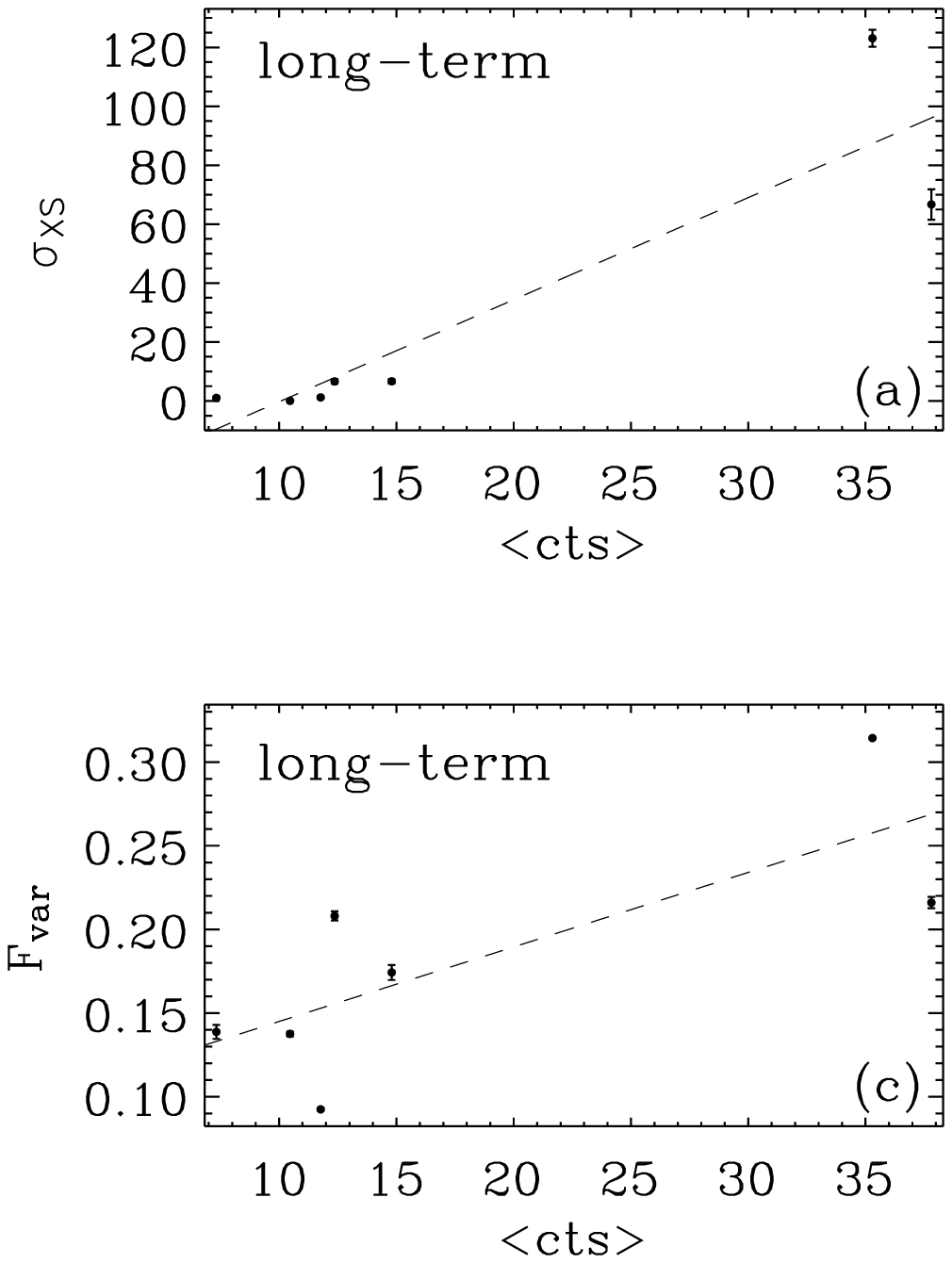}\includegraphics[bb=46 36 350 435,clip=,angle=0,width=8cm]{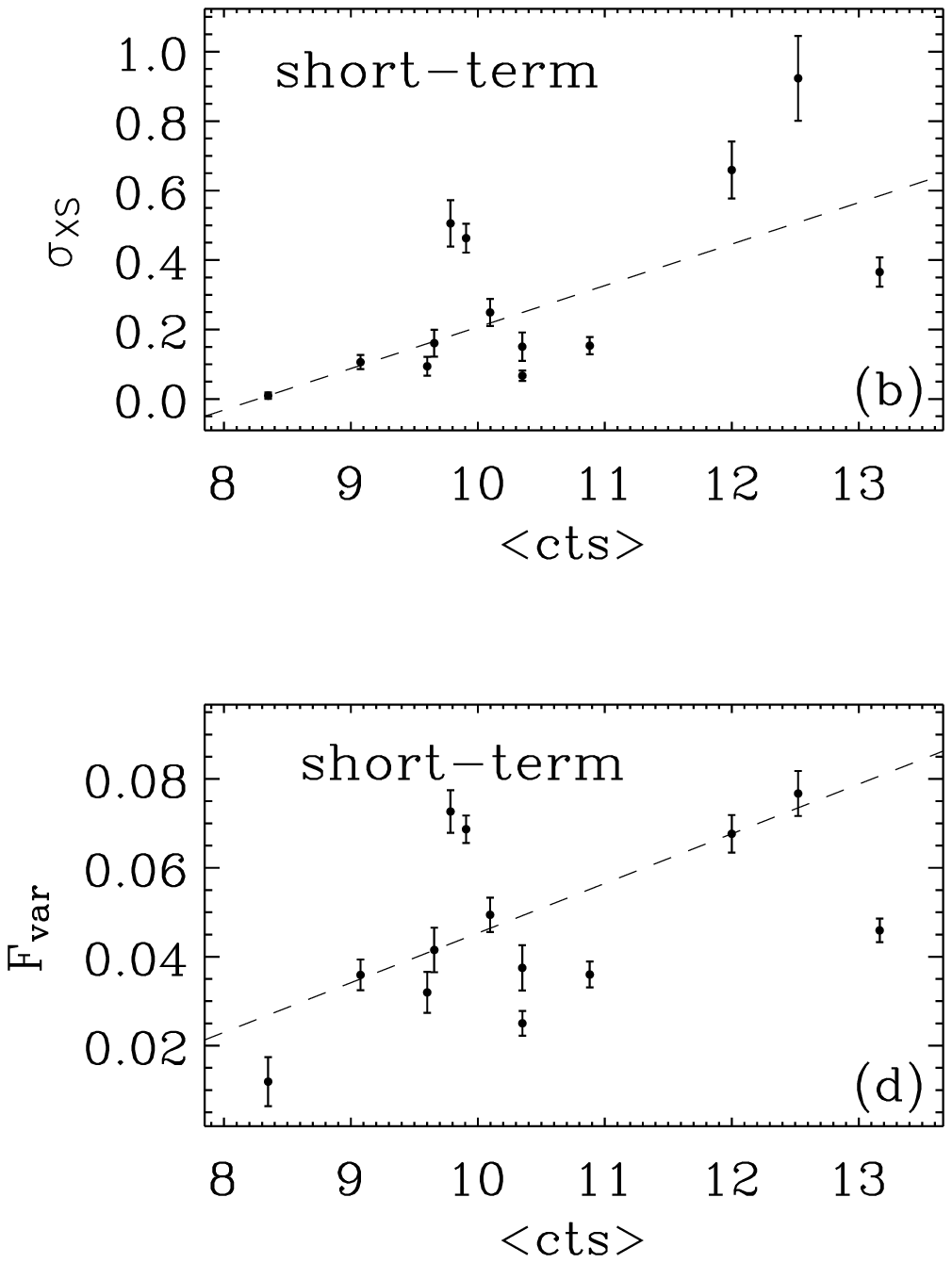}

\caption{Excess variance (top panels) and fractional variability parameter 
(bottom panels) plotted versus the mean count rate for the the long-term (a, c) and short-term (b, d) cases, respectively. The error-bars are computed according to Vaughan et al. 2003. The dashed lines represent the best fit values
obtained from a least square fitting procedure.} 
\label{figure:fig6}
\end{figure}

Figures~\ref{figure:fig6}a and \ref{figure:fig6}c, which show the 
excess variance and the
fractional variability parameter plotted against the mean count rate,
suggest the presence of a positive trend for both
quantities. This result is supported by a Spearman rank correlation
analysis, which yields chance probabilities of 0.2\% and 3.6\% for
$\sigma_{\rm XS}$ and $F_{\rm var}$, respectively. However, the small
number of data points available does not allow us to draw firm
conclusions. 

A similar analysis on shorter timescales has been carried out during
the denser monitoring campaign on May 1998  (Figures
\ref{figure:fig6}b and \ref{figure:fig6}d). Using time bins of 1000 s,
we divided a 14-day interval into 12 sub-intervals of duration of
$\sim$1 day containing 20 points each, and computed $\sigma_{\rm XS}$
and $F_{\rm var}$ for these sub-intervals.  Also in this case both
$\sigma_{\rm XS}$ and $F_{\rm var}$ seem to be positively correlated
with the mean count rate; however, only for $\sigma_{\rm XS}$ the
existence of the correlation is supported by a a Spearman rank
correlation analysis at a significant level ($P<5\%$),  whereas
the chance probability for a positive correlation between $F_{\rm var}$
and $ct$ is $P\sim 11\%$.
We caution that these results should be merely considered as
suggestive, due to their low statistical significance and due to the
unaccounted possible effects related to the red-noise nature of the 
variability (see Vaughan et al 2003 for a detailed discussion). A
deeper investigation of this issue can in principle be performed 
using extensive Monte Carlo simulations with specific assumptions
about the power spectrum model describing the X-ray variability. However,
this is beyond the scope of this paper, and, in any case, the 
outcome of these model-dependent tests would
be questionable,
since the characteristic of power spectral models for blazars are
poorly constrained.

\section{Correlated X-ray and TeV variability}

It is well known that Mrk~501 exhibits short-term, correlated flux 
variations at
X-ray and TeV wavelengths (Sambruna et al. 2000, Krawczynski et al.\ 2002; 
Katarzy\'nski et al. 2005). 
Here, we investigate whether the long-term X-ray and TeV
variations are correlated as well. We will also study
the correlated X-ray and TeV activities during the most prominent flares. 

\begin{figure}
\centering
\includegraphics[bb=35 30 355 296,clip=,angle=0,width=5.1cm]{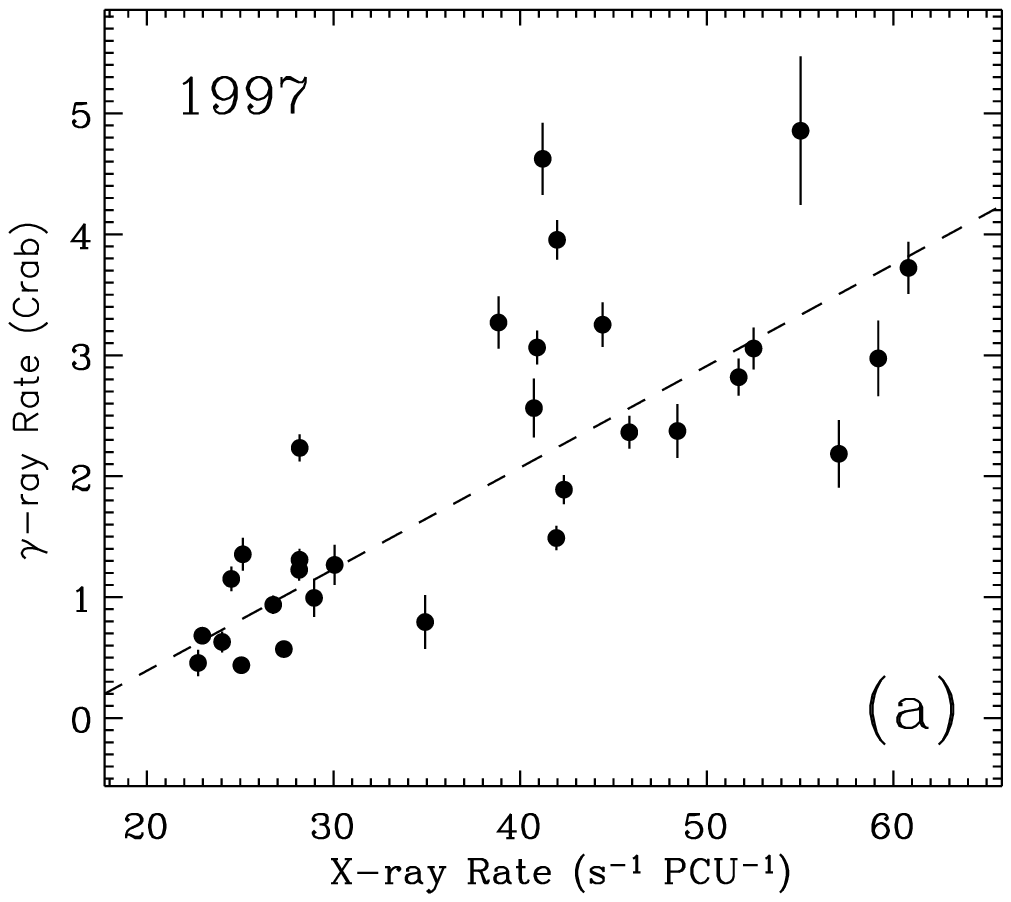}\includegraphics[bb=35 30 355 296,clip=,angle=0,width=5.1cm]{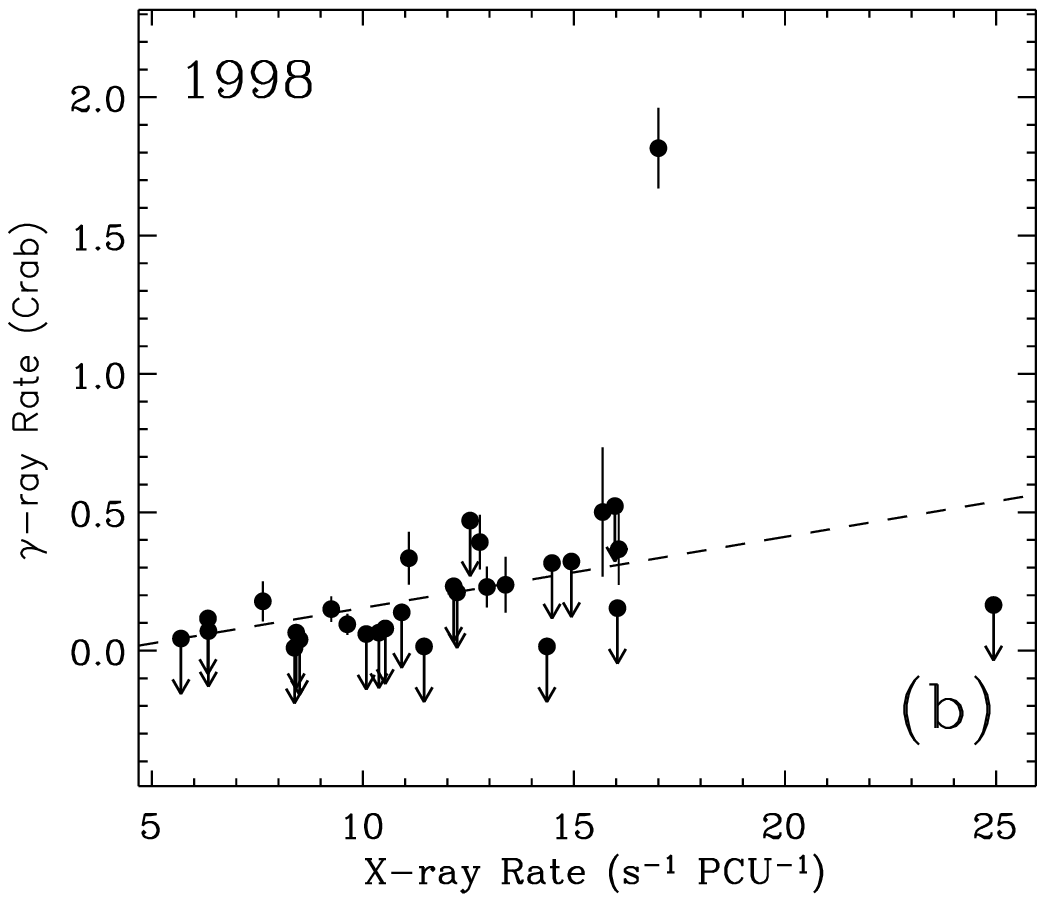}
\includegraphics[bb=35 30 355 296,clip=,angle=0,width=5.1cm]{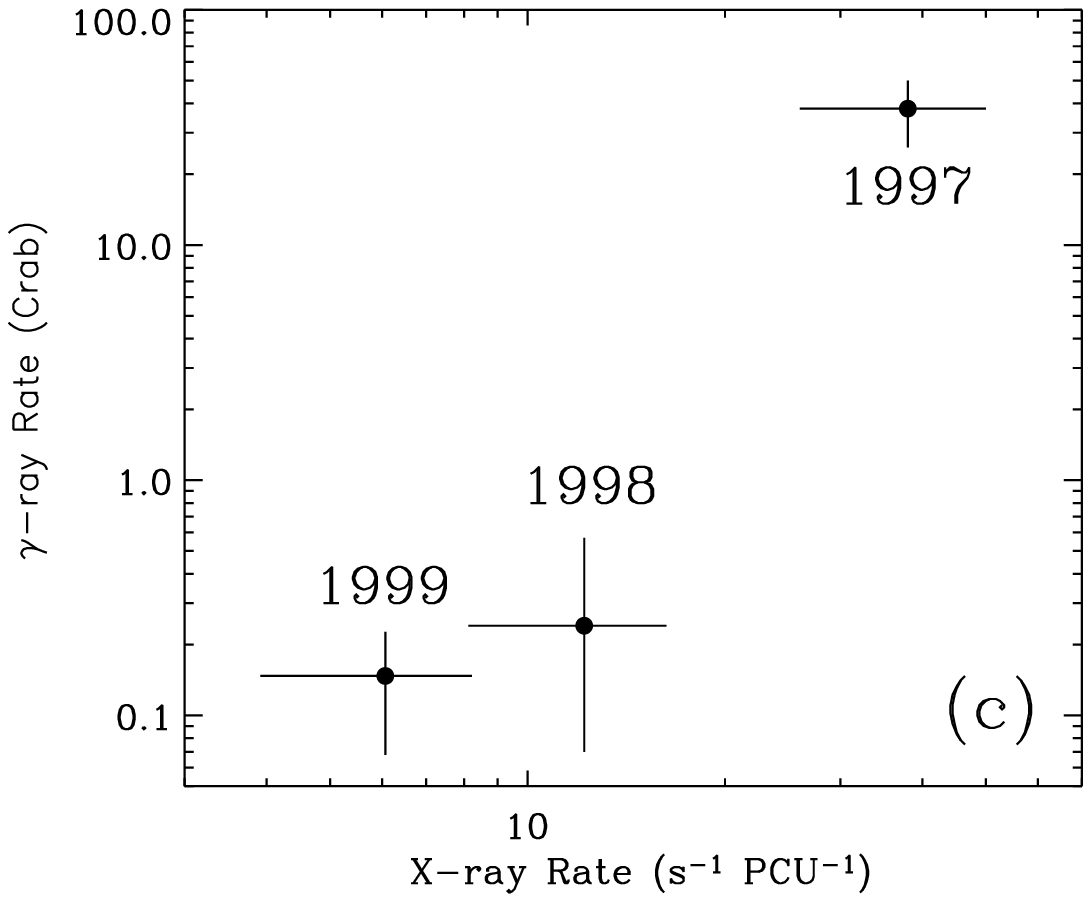}
\caption{$\gamma$-Ray count rate plotted against the 2-20 keV X-ray count rate
during the 1997 (a) and 1998 (b), respectively. Panel (c) shows the
mean values of the $\gamma$-ray fluxes (calculated using all the data points)
plotted against the X-ray count rates; the horizontal and vertical error-bars are the dispersion 
around the X-ray and $\gamma$-Ray mean values.}  
\label{figure:fig7}
\end{figure}

Figures \ref{figure:fig3}a--e show the 1997, 1998, 1999, and 2004 historical
X-ray light curves from \rxte, together with the historical TeV light
curves obtained by HEGRA (filled circles) and Whipple (open diamonds).
Most of the HEGRA data were originally discussed by Aharonian et
al. (2001), whereas most of the Whipple data were discussed by Quinn
et al. (1999). Despite the numerous gaps, correlated variability of
the flux at X-ray and TeV energies is present.  To illustrate
the presence of a positive trend during the long-term monitoring campaigns in 1997
and 1998, we plotted the TeV count rate against the
2--20 keV count rate (see Figure~\ref{figure:fig7} a and b). We
quantitatively tested the existence of  correlations, using
the methods described in $\S 3.1$. All methods consistently indicate the
presence of positive correlations in both monitorings. 
For example, in 1997 
we found $\rho=0.78$ ($P_{\rm Spearman}=3.2\times 10^{-7}$) and the slope
from a least square fit method is $b=0.084\pm0.002$,
whereas in 1998 we obtained
$\rho=0.54$ ($P_{\rm Spearman}=1.6\times 10^{-3}$) and $b=0.017\pm0.004$.
Due to the low-level of the TeV activity in 1998 coupled with the limited
sensitivity of the Cherenkov telescopes, many TeV fluxes are close to the detection
threshold and hence have large statistical errors, casting doubts on the 
statistical significance of the positive trend found. For this reason, we have
also used a more conservative approach considering as upper limits all the
data with flux $<$ 3$\sigma$. We used the statistical package ASURV
(Lavalley, Isobe \& Feigelson 1992) that is well suited for censored data,
obtaining $\rho=0.58$ and $P_{\rm Spearman}=1.6\times 10^{-3}$, which are
fully consistent with the previous results and confirm the existence of
a positive X-ray -- TeV correlation in 1998.

Figure~\ref{figure:fig7}c , which shows the TeV count rates, 
averaged over each contemporaneous
monitoring campaign, 
plotted versus the corresponding X-ray average count rates
suggests that a positive keV--TeV 
correlation exists on even longer timescales. The campaign in 2000 and 2004 were excluded, 
because in the first case
no coordinated TeV observations were performed, whereas in 2004 basically
all TeV measurements were upper limits.

\begin{figure}
\centering
\includegraphics[bb=55 60 295 320,clip=,angle=0,width=5.1cm]{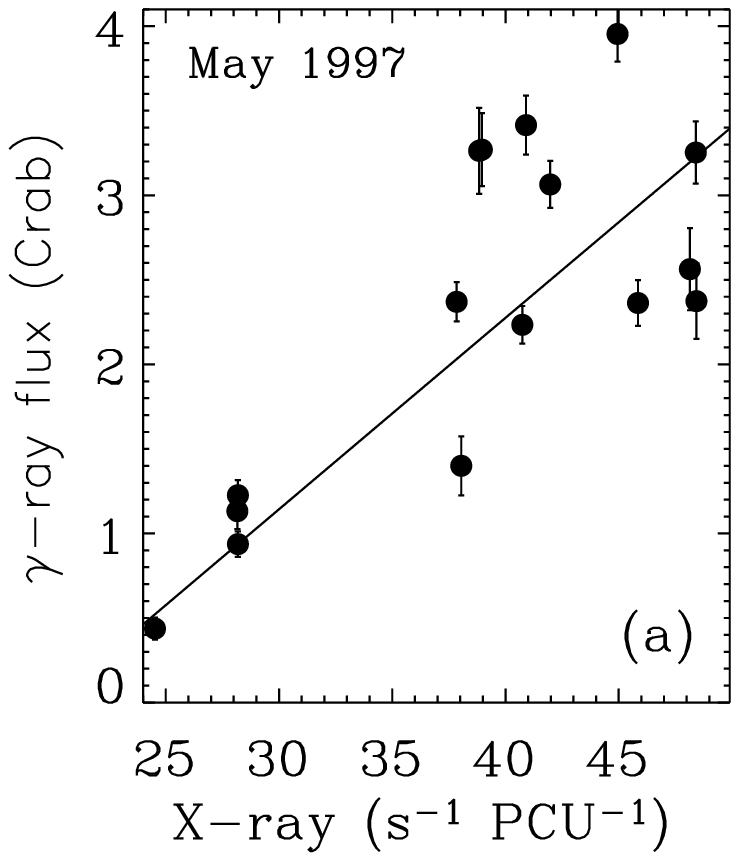}\includegraphics[bb=55 60 295 320,clip=,angle=0,width=5.1cm]{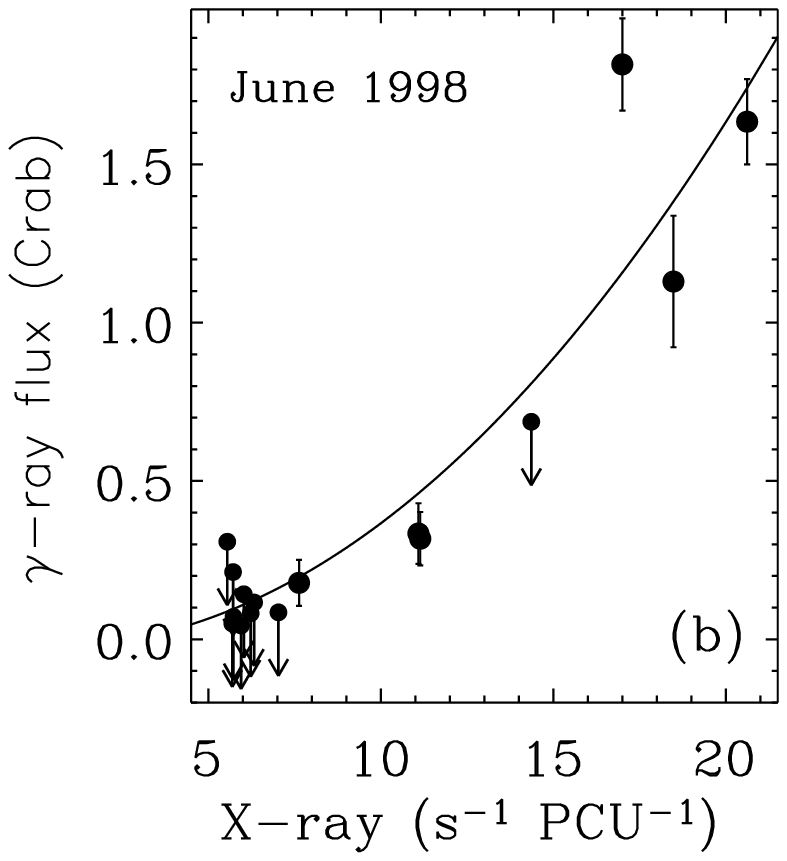}
\includegraphics[bb=55 60 295 320,clip=,angle=0,width=5.1cm]{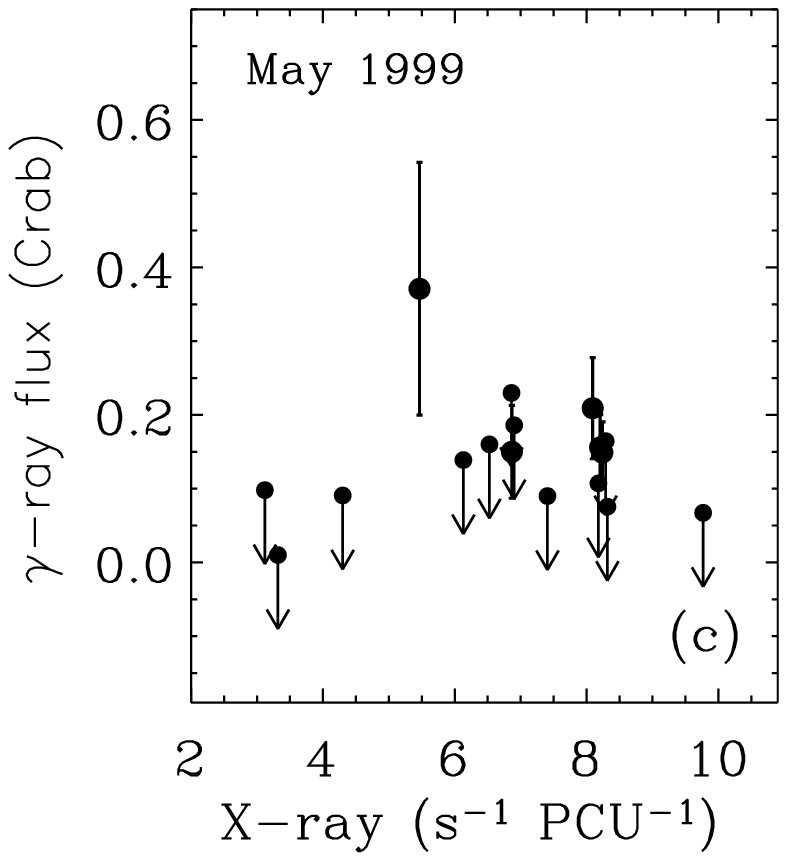}
\caption{TeV flux plotted versus the X-ray count rate 
for three representative flares, showing a linear correlation  $F_{\gamma}\propto 
F_{\rm X}^{0.99\pm0.01}$ (a), a nonlinear correlation $F_{\gamma}\propto 
F_{\rm X}^{2.07\pm0.36}$ (b), and no correlation at all (c).}  
\label{figure:fig8}
\end{figure}

Very few flares have sufficient coverage  to allow a meaningful
study of the X-ray/TeV correlation on shorter timescales. Nonetheless, it
is interesting to note the heterogeneous behavior shown by  different flares 
(Figure~\ref{figure:fig8}).  For example, in May 1997 (Fig.~\ref{figure:fig8}a),
a positive linear trend is observed: describing such trend
with a simple power law $F_{\gamma}\propto F_{\rm X}^{\alpha}$, we
find $\alpha=0.99\pm0.01$. On the other hand, in June 1998 (Fig.~\ref{figure:fig8}b),
a non-linear correlation ($\alpha=2.07\pm0.36$) seems to occur.
To test whether the latter data
can be instead described by a linear correlation, we fitted the 
data of June 1998 with $\alpha$ fixed to 1 and 2. The resulting fits 
indicate that
the non-linear case is preferable also on statistical grounds (the linear
case causes an increase of $\Delta\chi^2=7.1$ for the same number of
degrees of freedom). 
Unfortunately the paucity of the data and the large 
statistical errors hampers a more detailed analysis. 
Owing to large statistical errors, the data taken during 
May 1999 (Fig.~\ref{figure:fig8}c) do not show any evidence 
for a correlation.

A very interesting aspect of the relationship between the 
``simultaneously''\footnote{Fig.\ 3 uses X and TeV data points 
which have been taken within 8 hr from each other.}
measured X-ray and TeV fluxes is the substantial scatter observed in the
X-ray/TeV plane.
The 1997 observations show that for a given X-ray flux level,
the TeV fluxes can vary by factors up to $\sim$3, and for a given
TeV flux level, the X-ray fluxes can vary by factor up to $\sim$2 
(Fig.\ \ref{figure:fig7}a). In the 1998 data set there seems to 
be a linear correlation between the X-ray and TeV fluxes, 
but for a single flare, the TeV flux deviates by a factor of $\simeq$9 
from the flux predicted by the linear fit to the 
correlation (Fig.\ \ref{figure:fig7}b). Note that the data 
point lies 11.1 standard deviations above the linear fit 
taking into account both, the statistical uncertainties 
on the best-fit parameters and the statistical errors on the
X and TeV flux measurements. While it is tempting to refer to 
flares with relatively high TeV fluxes accompanied by low X-ray fluxes 
as ``orphan TeV flares'' and to flares with relatively high 
X-ray fluxes accompanied by low TeV fluxes as ``childless X-ray flares'',
this interpretation depends on how one defines a flare and the absence
of a flare in a certain frequency band. 

\begin{figure}
\centering
\includegraphics[bb=51 26 314 345,clip=,angle=0,width=7cm]{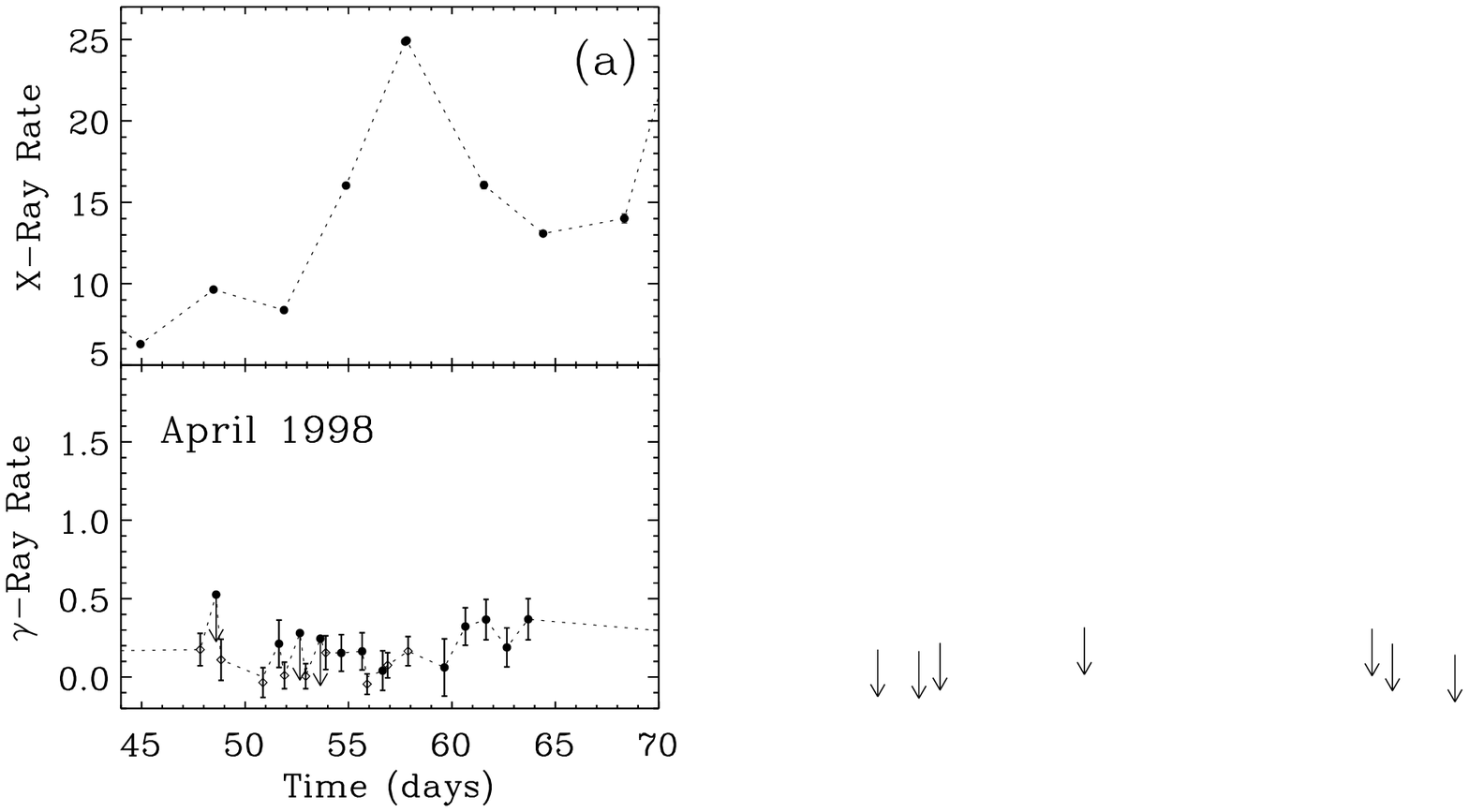}\includegraphics[bb=51 26 314 345,clip=,angle=0,width=7cm]{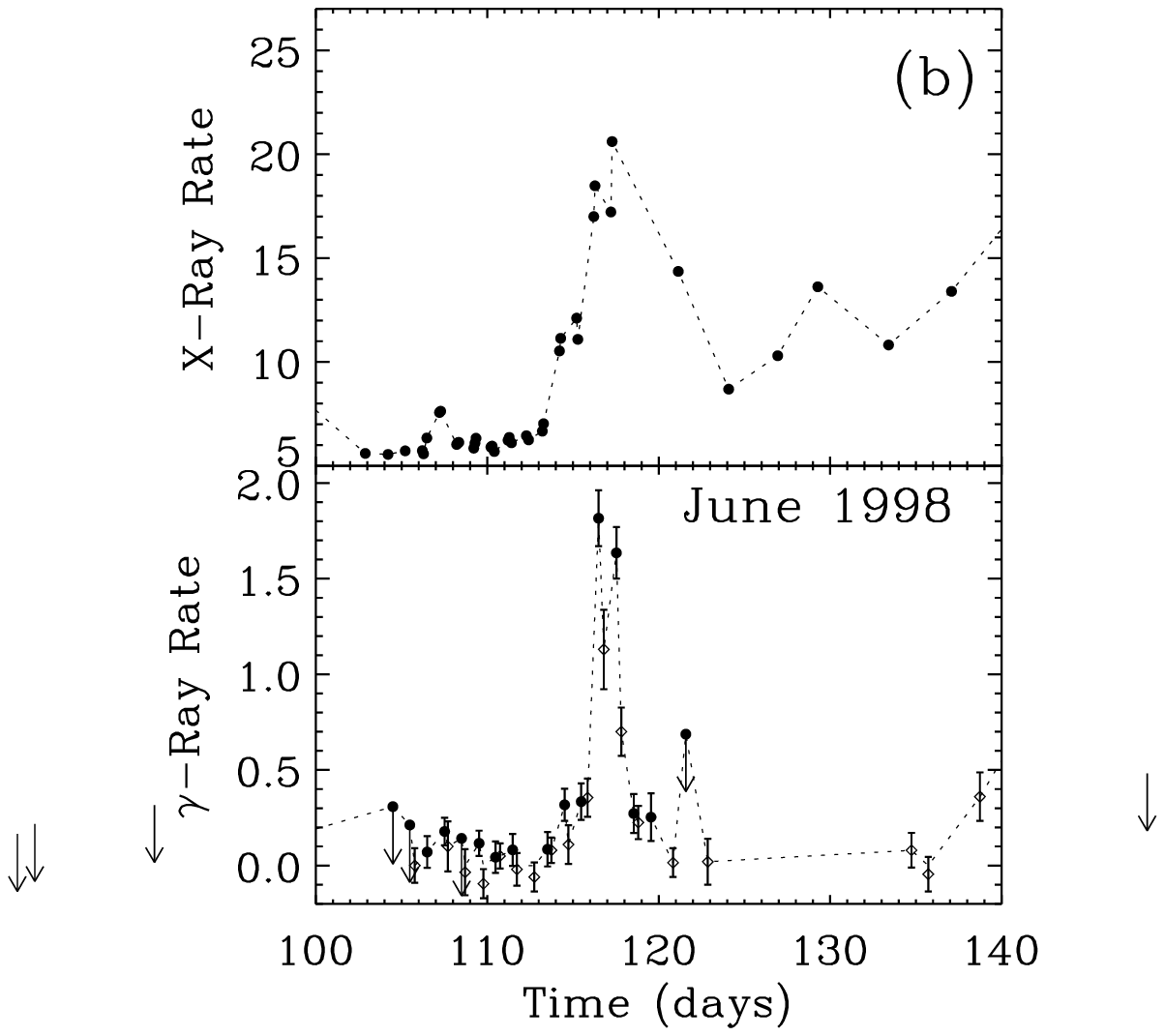}
\caption{$\gamma$-Ray and 2-20 keV X-ray light curves of a first X-ray flare apparently not accompanied by TeV activity (panel a) and a second X-ray flare of similar amplitude which
is accompanied by a contemporaneous prominent flare in the TeV energy band (panel b).
The panels are enlargements of the TeV and X-ray light curve shown in Fig. 3b.
}
\label{figure:fig9}
\end{figure}

A robust conclusion from the data sets discussed in this paper is that
the X/TeV correlation shows highly significant scatter, which suggests the
presence of uncorrelated activity in these two energy bands. 
To illustrate this point, we consider the enlargment of two flares
occurring less than three months apart during the 1998 coordinated
campaign (see Fig.~\ref{figure:fig9}). In both cases the X-ray count rate
increases by a factor $>$4 in less than one week. However, the associated
TeV activity, which on both occasions was well sampled with nearly one
observation per day, shows a strikingly different behavior. During the
X-ray flare on April 1998 the TeV level remains roughly constant around
the detection threshold. On the other hand, the X-ray flare in June 1998
is accompanied by a very prominent flare in the TeV energy band.

\section{Discussion} 
In this work we have undertaken a systematic study of the X-ray temporal and 
spectral
variability of the HBL Mrk~501 as seen with \rxte\ PCA, as well as of the 
long-term
correlation between the X-ray and TeV energy bands.

In the past years, several monitoring studies of the brightest blazars were 
carried out with different satellites; e.g., PKS~2155-304 was monitored 
with \rosat\ HRI (Brinkmann et al. 2000)
and later with \sax\ \citep{zhan99}.
Similarly Mrk~421 was monitored with \sax\ \citep{foss00}, 
and both sources along with Mrk~501 were also 
monitored with \asca\ \citep{tani01}.
However, the relatively short length (few weeks at most) of the light curves 
hampered the investigation of the long-term (months to years) temporal 
properties. Only the use of
\rxte, with its flexible observing schedule combined with a high sensitivity, 
has allowed one to probe the long-term variability in blazars. For example, 
using \rxte\ PCA data, Kataoka et al. (2001) studied the X-ray variability 
properties of Mrk~421, Mrk~501, and PKS~2155-304 on timescales ranging 
between $10^3$ and $10^8$ s, with structure functions
and power density spectra. 
Krawczynski et al. (2000, 2002) studied X/TeV observations of Mrk 501 
during the 1997 flaring epoch, focussing on the interpretation 
of the data in the framework of time-dependent 
Synchrotron Self-Compton models. More recently, Cui (2004) reported
the results from the flaring activity of Mrk~421 using archival \rxte\ data, 
and a similar investigation on Mrk~501 was carried out by Xue \& Cui (2005). 
Although 
the latter authors use data overlapping with ours, the aim of the X-ray 
temporal analysis is quite different: Xue \& Cui are mostly interested in 
the hierarchical nature of the flares and on the detection of the minimum  
characteristic timescale; on the other hand, our temporal analysis is aimed
at characterizing in a model-independent way the temporal and spectral 
variability of a jet-dominated source, with the final purpose
of shedding light on the jet contribution in non-jet-dominated AGN.

The main results from our model-independent analysis based on X-ray
colors and fractional variability can be summarized as follows: {\it on
long timescales} (i.e., from few to several months) 
the source spectrum hardens as the source brightens,
and the amplitude of variability progressively increases going from
the softest energy band (1--5 keV) to the hardest one (10--20
keV). These findings are typical for HBLs (e.g., Ulrich et al. 1997;
Aharonian et al. 2005).
Interestingly enough, the opposite spectral
and amplitude variability trends (e.g. Papadakis et al. 2002;
Markowitz \& Edelson 2001) are observed in radio-quiet AGN. In this case,
the spectrum becomes steeper as the flux increases, which can be
either explained in terms of a variation of the spectral slope,
predicted by thermal Comptonization models (e.g., Haardt \& Maraschi
1991), or in terms of a two-component model related to the accretion
process (Shih, Iwasawa \& Fabian 2002).

A direct consequence of the striking diversity between jet-dominated and
radio-quiet X-ray spectral properties, is that a model-independent 
analysis of the {\it long-term} X-ray variability can, in principle, put
constraints on the role played by the jet. 
This is particularly important for radio-loud
non-blazar objects, where a jet is certainly present but is not pointing
toward the observer. In this context, it is worth noticing that a similar 
analysis based on the  X-ray colors and fractional variability
was carried out by Gliozzi, Sambruna, \& Eracleous (2003) on two 
Broad Line Radio Galaxies (BLRG),
3C~390.3 and 3C~120, which, according to the Unification model, 
should be the same intrinsic objects as blazars, but seen at a larger 
viewing angle. 
Interestingly enough the BLRGs show a trend of decreasing
$F_{\rm var}$ with increasing energy, and a trend of decreasing hardness 
ratio with increasing count rate. This suggests the existence
of a different process causing X-ray variability in the two radio-loud 
classes of AGN, and that the X-ray radiation in BLRGs is unlikely dominated 
by the jet emission.
However, before assuming as a universal, genuine signature of the jet
the X-ray spectral variability observed in Mrk~501, the same 
model-independent analysis should be carried out on other jet-dominated 
objects. For example, objects belonging to the Low-energy peaked BL Lacs
(LBL) and intermediate BL Lacs have been known to show short-term spectral and
temporal properties different from HBL (e.g., Urry et al. 1996;
Tanihata et al. 2000). We plan
to undertake this analysis in a future work, using \rxte\ archival data.

It is worth pointing out that the previous findings refer to long 
timescales (from few to several months). When shorter timescales are considered, 
for example in the case of single flares, the typical HBL spectral and 
variability trend (i.e., harder spectrum--when--brighter and variability 
amplitude larger at harder energies) is still observed in most of the cases,
but not in all of them. The most noticeable exception is 
the event observed in July 1997 (Fig. 4e, 5e), which shows the opposite 
temporal and spectral trends. Below we discuss this peculiar event in more detail.

As already mentioned, X-ray emission from HBL like Mkn 501 is almost
unanimously attributed to synchrotron emission. In this framework, the study
of the variability displayed by the X-ray continuum offers the
opportunity to probe the dynamics of the underlying population of
relativistic electrons. In this context the general variability trend
($F_{\rm var,hard} > F_{\rm var,soft}$) is readily explained by the
fact that the hard X-rays are produced by more energetic particles,
that are characterized by shorter cooling timescales causing the
higher variability amplitude observed at hard X-rays.  Analogously,
the general spectral trend (source harder when brighter), including
the presence of hysteresis loops in the $HR - ct$ plane, carries
important information on the acceleration and cooling processes acting
on the emitting electrons. Indeed these characteristics are often
explained in terms of different ratios between the acceleration and
the cooling timescales of the relativistic electrons (Kirk et
al. 1998; Ravasio et al. 2004).
In this regards, particularly intriguing is the event observed in July
1997. This event, which comprises the decay phase of a first flare
followed by the rising phase of a second flare, has been already
investigated in details by Lamer \& Wagner (1998). The main result of
the above study, in addition to the alreay mentioned anti-correlation
$HR\propto 1/ct$, is that the soft and hard X-ray light curves  
vary in a completely independent way. The main
conclusion derived by Lamer \& Wagner was that at least two emission
components are necessary to explain this unusual behavior. However, no
explanation on the physical nature of these components was proposed,
apart from the implicit assumption that part of the X-rays are
produced by synchrotron emission from the low-energy tail of the
electron distribution. 

We have re-analyzed the \rxte\ PCA data of July 1997; the details of this
analysis are reported in Appendix 2. The first important point to consider
is that this event (observed with 10 pointings over 5 consecutive days) is
made up of several sub-flares on timescales of few hundreds of
seconds (see Fig. 10). We investigated the trend of the fractional variability
versus the energy for the longest individual pointings (duration $> 2500$ s)
and found that $F_{\rm var,soft}$ is never significantly larger than 
$F_{\rm var,hard}$. In fact, a positive trend is found in the $F_{\rm var} - E$
plane, when the source shows significant variability (on July 13 and 16; see 
Fig. 11), whereas no trend is observed when the short-term light curve is
roughly constant (July 14 and 15). As a consequence, the ``anomalous" inverse
correlation obtained combining all the data of July 1997 (Fig. 5e) is likely
to be a spurious result. 

For the same long pointings, we plotted the 
X-ray hardness ratio versus the count rate (Fig. 11, panels of the second column) and found
no evidence for any inverse correlation on short timescales; a ``normal"
positive trend is observed in all the pointings, except for July 14, when a
constant trend is present. However, combining the individual pointings, an
inverse correlation is observed,
i.e., a brightening of the source accompanied by spectral
softening. To investigate further this issue, we performed a 
time-resolved spectral analysis on the individual, long exposures, using the
most recent PCA background and calibrations files. The results of this analysis,
summarized in Table 3, reveal that all the spectra are well described by a broken
power-law model with the same spectral parameters ($\Gamma_1\sim 1.7-1.75$,
$E_{\rm br}\sim 6.5-7$ keV,  $\Gamma_2\sim 1.85-1.90$), except for the observation on July 16,
which is characterized by the highest count rate. On that occasion,
a significant steepening of the hard photon index is observed ($\Gamma_2\sim 2$).
In the framework where the X-rays are produced via synchrotron emission, $\Gamma_2\sim 2$
represent the location of the synchrotron peak in the $\nu F_\nu {\rm ~vs.~} \nu$ plot.
In this context, the time-resolved spectral analysis seems to reveal an unusual behavior,
suggesting the presence of an anti-correlation between the synchrotron peak and the
source luminosity. However, a better coverage (i.e., with uninterrupted observations
and higher quality spectra) of Mrk~501 during its highest activity level is necessary
to confirm this result.

In our model-independent analysis of the X-ray variability of Mrk~501
we have also investigated the possible presence of a correlation between the
fractional variability $F_{\rm var}$ and the mean count rate, and 
between the excess variance $\sigma_{\rm XS}$ and the mean count rate.
 Our analysis is motivated by recent work from 
Uttley and
collaborators (Uttley et al. 2005 and references therein) that revealed, 
both for radio-quiet AGN and GBHs, the existence of a positive linear 
correlation  between the amplitude of the X-ray variability (as measured by 
the {\it non-normalized} excess variance) and the X-ray flux. Note that, when the 
excess variance
is normalized to the mean count rate (as in the case of $F_{\rm var}$), the positive
correlation vanishes. Uttley et al. (2005) show that the presence of this
rms-flux correlation sets tight constraints on variability models, ruling out
all intrinsically linear models. Recently, this analysis has been 
applied by Zhang et al. (2005) to the jet-dominated source PKS~2155-304, 
using \xmm\ data. Similarly to radio-quiet AGN, PKS~2155-304 shows, on short 
timescales, a positive correlation between $\sigma_{\rm XS}$ and the mean 
count rate, however, it also shows a negative correlation between 
$F_{\rm var}$ and the mean count rate. 
Our analysis, carried out using homogeneously sampled intervals of 
$\sim$2 weeks and spanning a time period of 3--4 years,
suggests the presence of a positive correlation with the mean count rate
of  both $F_{\rm var}$
and $\sigma_{\rm XS}$. We caution,
however, that, due to the paucity of data points
and the unaccounted possible effects of the red-noise nature of the variability, 
this result should be considered
merely as suggestive. Long uninterrupted yearly-long monitoring campaigns 
of blazars (similar to those carried out for radio-quiet AGN)
are necessary to derive firmer conclusions on this issue.

The second important goal of our analysis is to investigate the
long-term correlation between the X-ray and TeV bands. It is well
known that in these sources X-rays and TeV emissions are positively
correlated on a wide range of different timescales, from hours (e.g.,
Maraschi et al. 1999; Fossati et al. 2004) to weeks-months (e.g.,
Krawczynski et al. 2000; Blazejowski et al. 2005).  As expected, we
confirm that Mrk~501 shows a positive correlation between the X-ray
and TeV emission. 
An interesting feature, revealed by the analysis of correlated 
X-ray and TeV activities, is that the X/TeV correlation exhibits
a considerable scatter: a range of X-ray fluxes is observed for a given TeV
flux and a range of TeV fluxes is observed for a certain X-ray flux.
The scatter may have several explanations. For example, if flares
are shorter or longer in the TeV band than in the X-ray band
(owing for example to different radiative cooling times of the particles
responsible for the TeV and X-ray emission), the flux ratio changes 
during a flare. If physical parameters characterizing the emitting 
plasma (e.g. the magnetic field, or the size or Doppler factor of 
the emitting plasma) change over time, the X/TeV flux ratio would change
during a flare or from flare to flare. Furthermore, orphan TeV 
flares or X-ray flares without TeV counterpart may complicate the picture.

The presence of a positive X-ray -- TeV correlation is, at first
order, a prediction of SSC models, where the X-ray emission is
produced by the synchrotron radiation from highly relativistic
electrons in the jet, which are also responsible for the TeV emission
through the inverse-Compton scattering process off the synchrotron
photons (that, for high energies involved here, occurs in the
Klein-Nishina regime). Alternatively, this correlation can be
explained by hadronic models, where the X-ray emission is still
attributed to synchrotron radiation from relativistic electron in the
jet, but the TeV emission is supposed to be produced by particles
resulting from cascades triggered by relativistic protons (e.g.,
M\"ucke et al. 2003) or by the protons themselves through synchrotron
emission. However, it is worth noting that, for the typical duration
of the flares analized in this work ($\Delta t\sim 10$ days), the
emission cannot be attributed to a single, moving blob, as usually
assumed in the simplest versions of these models, usually applied to
shorter events (Katarzynski et al., in prep). Indeed, for typical
Lorentz factor $\Gamma =10$, the blob would travel for a distance $d>
1 $ pc, implying a huge expansion and thus a strong variation of the
physical parameters and of the timescale of the variations, in
contrast with the substantial stability of the observed emission. Thus
the observed flares are probably composed by several subflares
produced in different regions of the flow. For this reason the
presence of the large scattering observed in the X-ray -- TeV
correlation on long timescale is not surprising.

However, other observational results are starting to put stronger
constraints on the models.  For example, it is difficult to explain
the existence of X-ray flares not accompanied by TeV flares, as the
ones possibly observed in Mrk~501, and of ``orphan" TeV flares, as
observed in 1ES1959+650 (Krawczynski et al.  2004; but see the
hadronic scenario proposed by Boettcher 2005).  Further constraints on
theoretical models can be placed by the analysis of individual flares,
if a non-linear correlation (i.e., $F_{\gamma}\propto F_{\rm
X}^{\alpha}$ with $\alpha>1$) is detected. In particular, as discussed
in detail by Katarzy\'nski et al. (2005), current SSC models face a
difficult challenge in explaining the presence of a non-linear
correlation during the decay phase of the flare. Unfortunately, the
sparse coverage of the decay phase in the flares of Mrk~501 monitored
with \rxte\ PCA does not allow one to investigate further this
important issue.

Higher quality simultaneous X-ray -- TeV observations both on short (i.e.,
individual flares) and long timescales are critically
necessary to establish firmly the nature of this correlation
and the presence (or lack) of genuine orphan flares, which
in turn will possibly allow to distinguish between the competing theoretical models.

\section{Summary and Conclusions}
 
We presented the results of a temporal and spectral analysis of \rxte\
observations at X-rays of the nearby TeV blazar Mrk~501, performed
between 1997 and 2004. The goal of our ongoing study is to
characterize the long-term X-ray flux and spectral variability of this
source, and compare it to the TeV long-term light curve.
The main findings and conclusions can be
summarized as follows:
\begin{itemize}

\item The first important result of the variability analysis of Mrk~501
is that the source has a very strong spectral variability and that on
long-term timescales the X-ray colors are positively correlated with the 
count rate: the higher the count rate, the harder the spectrum. 
A similar spectral behavior is found in most of the individual flares,
with a noticeable exception in July 1997.

\item Other important findings from the temporal analysis are obtained by
applying the fractional variability analysis to energy-selected light
curves. We found that the fractional variability amplitude
seems to be positively correlated with the energy band.

\item If the above spectral variability trends 
($HR \propto ct$,  $F_{\rm var} \propto E$),
which seem to be typical for the long-term behavior of HBL objects,
are confirmed also for the other blazar classes, they will provide a
model-independent way to single out the jet contribution in radio-loud
AGN and possibly in GBHs in the low-hard state.

\item Comparing the historical RXTE, HEGRA, and Whipple light curves, we find
hints for correlated X-ray and TeV flux variations on {\it long}
timescales (weeks to months). This confirms and extends the observed
correlated variability at these two wavelengths previously observed on
shorter timescales. The origin of the TeV flux is still an open question.

\item Individual flares show a heterogeneous behavior when the TeV flux
is plotted versus the X-ray flux: Some flares show a linear TeV -- X-ray
correlation, and some show a non-linear correlation. 
Furthermore, the X/TeV flux correlation shows considerable scatter 
with similar X-ray (TeV) fluxes being accompanied by more than a factor of 2
different TeV (X-ray) fluxes.
\end{itemize}

\begin{acknowledgements} 
MG acknowledges support from NASA LTSA grant NAG5--10708. RMS was
funded by an NSF CAREER award and from the Clare Boothe Luce Program
of the Henry Luce Foundation while at George Mason University. 
F.T. acknowledges support from COFIN grant 2004023189$\_$005. 
HK and IJ acknowledge support by the RXTE Guest Investigator Program
under NASA grant NNG04GQ26G.
This
research has made use of data obtained through the High Energy
Astrophysics Science Archive Research Center Online Service, provided
by the NASA/Goddard Space Flight Center.
\end{acknowledgements}

\begin{table} 
\caption{Correlation hardness ratio vs. Count rate}
\begin{center}
\begin{tabular}{lrllrllrlll} 
\hline        
\hline
\noalign{\smallskip}        
Long-term & $r$ & $(P_{\rm Pearson})$ && $\rho$  & $(P_{\rm Spearman})$ && $\tau$  & $(P_{\rm Kendall})$ && $b^{\mathrm{a}}$ \\
\noalign{\smallskip}  
\hline  
\noalign{\smallskip}
1997 & $0.80$ &$(< 1\times10^{-5})$ && 0.56 &($< 1\times10^{-5}$) && 0.71 &($< 1\times10^{-5})$ && 0.015($\pm 0.003$)\\
\noalign{\smallskip}
\hline
 \noalign{\smallskip}
1998 & $0.39$ &$(< 1\times10^{-5})$ && 0.31 &($< 1\times10^{-5}$) && 0.24 &($< 1\times10^{-5})$ && 0.020($\pm 0.001$)\\
\noalign{\smallskip}
\hline
  \noalign{\smallskip}
2000 & $0.87$ &$(< 1\times10^{-5})$ && 0.45 &($5.4\times10^{-4}$) && 0.34 &($2.9\times10^{-4})$ && 0.134($\pm 0.002$)\\
\noalign{\smallskip}
\hline
\hline
\noalign{\smallskip}        
Flare & $r$ & $(P_{\rm Pearson})$ && $\rho$  & $(P_{\rm Spearman})$ && $\tau$  & $(P_{\rm Kendall})$ && \\
\noalign{\smallskip}  
\hline  
\noalign{\smallskip}
Apr 97 & $0.94$ &$(< 1\times10^{-5})$ && 0.96 &($< 1\times10^{-5}$) && 0.87 &($< 1\times10^{-5})$ && \\    
\noalign{\smallskip}                 
\hline  
\noalign{\smallskip}
Jul 97 & $-0.68$ &$(4.0\times10^{-3})$ && -0.45 &($1.5\times10^{-2}$) && -0.71 &($1.9\times10^{-3})$ && \\    
\noalign{\smallskip}
\hline               
\noalign{\smallskip}
May 98 & $0.70$ &$(5.6\times10^{-5})$ && 0.57 &($2.1\times10^{-3}$) && 0.43 &($1.6\times10^{-3})$ && \\                     
\noalign{\smallskip}
\hline  
\noalign{\smallskip}
Jun 98 & $0.90$ &$(< 1\times10^{-5})$ && 0.64 &($4.2\times10^{-5}$) && 0.49 &($5.2\times10^{-5})$ && \\                     
\noalign{\smallskip}
\hline  
\noalign{\smallskip}
May 99 & $0.53$ &$(1.2\times10^{-4})$ && 0.41 &($3.1\times10^{-3}$) && 0.30 &($2.8\times10^{-3})$ && \\                     
\noalign{\smallskip}
\hline  
\noalign{\smallskip}
Jun 04 & $0.15$ &$(5.7\times10^{-1})$ && 0.36 &($1.5\times10^{-1}$) && 0.25 &($1.6\times10^{-1})$ && \\                     
\noalign{\smallskip}
\hline  
\end{tabular}
\end{center}

$^{\mathrm{a}}$ Best fit value and 1 $\sigma$ error of the slope $b$
(with $y=a+bx$) obtained from a least square fit.
\label{tab1}
\end{table}

\begin{table} 
\caption{Correlation Fractional Variability vs. Energy}
\begin{center}
\begin{tabular}{lrllrlll} 
\hline        
\hline
\noalign{\smallskip}        
Long-term & $\rho$  & $(P_{\rm Spearman})$ && $\tau$  & $(P_{\rm Kendall})$ && $b^{\mathrm{a}}$ \\
\noalign{\smallskip}  
\hline  
\noalign{\smallskip}
1997$^{\mathrm{b}}$ & 0.99 &($< 1\times10^{-5}$) && 0.96 &($1.20\times10^{-4})$ && $(3.9 \pm 0.2)\times10^{-3}$\\
\noalign{\smallskip}
\hline
 \noalign{\smallskip}
1997 & 1.00 &($< 1\times10^{-5}$) && 1.00 &($1.17\times10^{-1})$ && $(4.4 \pm 0.2)\times10^{-3}$\\
\noalign{\smallskip}
\hline
 \noalign{\smallskip}
1998 &  1.00 &($< 1\times10^{-5}$) && 1.00 &($1.17\times10^{-1})$ && $(6.3 \pm 0.3)\times10^{-3}$\\
\noalign{\smallskip}
\hline
  \noalign{\smallskip}
2000 &  1.00 &($< 1\times10^{-5}$) && 1.00 &($1.17\times10^{-1})$ && $(3.2 \pm 0.1)\times10^{-2}$\\
\noalign{\smallskip}
\hline
\hline
\noalign{\smallskip}        
Flare &  $\rho$  & $(P_{\rm Spearman})$ && $\tau$  & $(P_{\rm Kendall})$ &&  $b^{\mathrm{a}}$ \\
\noalign{\smallskip}  
\hline  
\noalign{\smallskip}
Apr 97 & 1.00 &($< 1\times10^{-5}$) && 1.00 &($1.17\times10^{-1})$ && $(9.5 \pm 0.2)\times10^{-3}$ \\    
\noalign{\smallskip}                 
\hline  
\noalign{\smallskip}
Jul 97 & -1.00 &($< 1\times10^{-5}$) && -1.00 &($1.17\times10^{-1})$ && $(-3.1 \pm 0.3)\times10^{-3}$ \\    
\noalign{\smallskip}
\hline               
\noalign{\smallskip}
May 98 & 1.00 &($< 1\times10^{-5}$) && 1.00 &($1.17\times10^{-1})$ && $(3.1 \pm 0.4)\times10^{-3}$\\                     
\noalign{\smallskip}
\hline  
\noalign{\smallskip}
Jun 98 & 1.00 &($< 1\times10^{-5}$) && 1.00 &($1.17\times10^{-1})$ && $(2.0 \pm 0.1)\times10^{-2}$\\                     
\noalign{\smallskip}
\hline  
\noalign{\smallskip}
May 99 & 0.50 &($ 6.67\times10^{-1}$) && 0.33 &($6.02\times10^{-1})$ && $(7.0 \pm 0.2)\times10^{-3}$\\                     
\noalign{\smallskip}
\hline  
\noalign{\smallskip}
Jun 04 & -0.50 &($ 6.67\times10^{-1}$) && -0.33 &($6.02\times10^{-1})$ && $(1.2 \pm 1.1)\times10^{-3}$\\                     
\noalign{\smallskip}
\hline  
\end{tabular}
\end{center}

$^{\mathrm{a}}$ Best fit value and 1 $\sigma$ error of the slope $b$
(with $y=a+bx$) obtained from a least square fit.\\
$^{\mathrm{b}}$ Results obtained using ten narrower energy bands instead of three.
\label{tab1}
\end{table}

\clearpage

%

\clearpage
 
\appendix
\section{Time-averaged spectral analysis}
Since the data consist of short snapshots spanning a long temporal baseline, 
showing
the presence of pronounced temporal and spectral variability, a time-averaged spectral
analysis might not be the most appropriate tool for probe the physical properties of
Mrk~501. Nonetheless, useful constraints on physical models can be inferred from the
spectral analysis.
The main results from the spectral analysis of the 1998--2000 \rxte\ campaigns
are summarized below (the results of the 1997 campaigns have been already
discussed at length in previous works).

\noindent {\em The 1998 Campaign}\\
The time-averaged X-ray spectrum for the 1998 data is best fitted by a broken power-law
model with Galactic absorption and photon indices $\Gamma_{\rm low}=2.04 \pm 0.01$ 
and  $\Gamma_{\rm high}=2.15 \pm 0.01$, with a break energy of $E_{\rm break}= 6.3\pm 0.3$. 
The corresponding 2--10 keV flux is  F$_{2-10~keV}=1.0 \times 10^{-10}$ \flux.  
The HEXTE spectrum in the range of 20--60 keV can be described by a power-law model with 
$\Gamma=2.2\pm0.2$.
Spectral fits performed on the individual observations with a single power law show
that the photon index  ranges between $\sim 1.9$ and $\sim 2.4$, with steeper values
generally corresponding to lower fluxes. However, when the photon indices are plotted
against the flux, a large scattering is found. 

\noindent {\em The 1999 Campaign}\\
The average PCA X-ray spectrum for the 1999 data can be well described by a power-law fit with parameter values of $\Gamma=2.32\pm 0.01$.
The average HEXTE spectrum for the 1999 data is poorly constrained, but consistent
within the errors with the PCA spectrum:  $\Gamma=
2.7^{+1.6}_{-0.9}$.

\noindent {\em The 2000 Campaign}\\
The PCA spectra of each individual observation was fitted with either a power law or broken power law  absorbed by Galactic N$_{H}$=1.7e+20. 
A broken power-law model 
provides a significantly better fit in 40\% of the spectra, according to an F-test.
This model always yields a {\it concave} continuum, with $\Gamma_{\rm low}\sim 2$,
$\Gamma_{\rm high}\sim 1$ and break energy around 9--10 keV. 
For a single power-law model fits, spectral changes up to $\delta\Gamma=0.5$ are present
for modest (20\%) flux changes, with an erratic behavior. 
The time-averaged PCA spectrum was fit with a power-law model from 2--20 keV, 
yielding a photon index of  $\Gamma$=$2.15\pm{0.01}$. This value is consistent
within the errors with the one obtained from the time-averaged HEXTE spectrum (20--60 keV),
 $\Gamma$=$2.4^{+0.5}_{-0.6}$.

\section{The ``anomalous" event in July 1997}
Since the event in July 1997 is the only occasion where Mrk~501 clearly shows
an anti-correlation in the $HR - ct$ and $F_{\rm var} - E$ plots, we have decided
to carry out additional temporal and spectral analyses using smaller time-bins.

Fig. 10 shows the 2--20 keV PCA light curve of Mrk~501
during July 1997 with tbin=240s. The source was monitored for 5 consecutive days
with two pointings per day. With the exception of the first observation on July 12
where only short pointings ($< 1500$ s) were performed, all the other observations 
are characterized by a short exposure followed by a relatively long ($\sim 3000$ s) exposure. 
We will focus our analysis on these longer
pointings, since they allow a more detailed study of the temporal and spectral properties.
The first important feature to notice in Fig. 10 is the presence of several sub-flares
on timescales of few hundreds of seconds. This is better illustrated by the first column
of Fig. 11, which reveals strong variability on short timescales during July 13 and 16
and weaker amplitude variations on July 14 and 15.
The second and third columns represent the $HR -ct$ and $F_{\rm var} - E$ plots, respectively.
The main findings of this model-independent analysis can be summarized as follows:\\
1) When significant variability is observed (July 13 and 16), the spectral and 
temporal variability follow the general trend, which is observed on longer timescales and 
on most of the flares, with the source spectrum that hardens as the source brightens, and 
the amplitude of variability progressively increasing from the softest energy band to the 
hardest one.\\
2) When no significant X-ray variability is observed (July 14 and 15), the spectral variability
is less pronounced with a weak positive correlation in the $HR -ct$ plot and no correlation
at all in the $F_{\rm var} - E$ plane. Note that, when the light curves are nearly constant,
the variability is dominated by statistical uncertainties hampering the determination
of $F_{\rm var}$. This explains the presence of two data-points only in the $F_{\rm var} - E$
plots during July 14 and 15.

Importantly, the ``anomalous" variability trend (as shown in Fig. 5e) is 
never observed in any of the individual exposures. 
The fact that in each individual observation (even during the short 
pointings studied using time-bin=100s, and not shown in Fig. 11)
$F_{\rm var,soft}$ is never larger than  $F_{\rm var,hard}$ 
suggests that the inverse correlation $F_{\rm var} \propto 1/E$ found considering the
entire interval 12--16 July 1997 is probably an artifact, due to the combination
of several heterogeneous sub-flares poorly sampled. 

Also the hardness ratio versus the count rate shows a positive (or, at the very least, 
constant) trend during each individual observation (see the second column panels of Fig. 11). 
However, on longer timescales
(i.e., combining the individual observations) an opposite negative trend is found.
To investigate this issue in greater detail, we have carried out a time-resolved
spectral analysis on the longest individual pointings. All the spectra are well fitted by
a broken power law absorbed by Galactic absorption. Interestingly, while the spectra on July 13, 14, and 15 share the same spectral parameters within the errors, on July 16, when the source was at its brightest level, a significant 
($> 3 \sigma$) steepening of the hard photon index is observed.
These results are summarized in Table 3, where the errors quoted are the 90\% confidence levels. 

\begin{figure}
\includegraphics[bb=50 30 520 290,clip=,angle=0,width=16.cm]{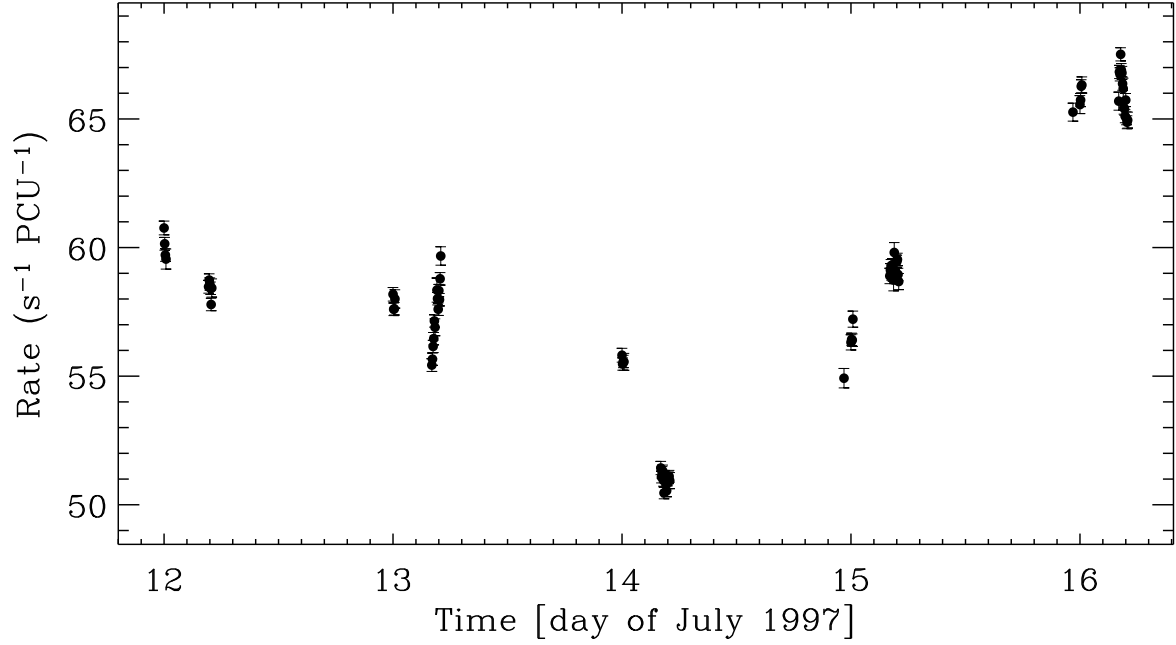}
\caption{\rxte\ PCA light curve of Mrk~501 during July 1997 1.
Time bins are 240 s.} 
\label{figure:fig10}
\end{figure}

\begin{figure}
\includegraphics[bb=60 130 569 691,clip=,angle=0,width=16.cm]{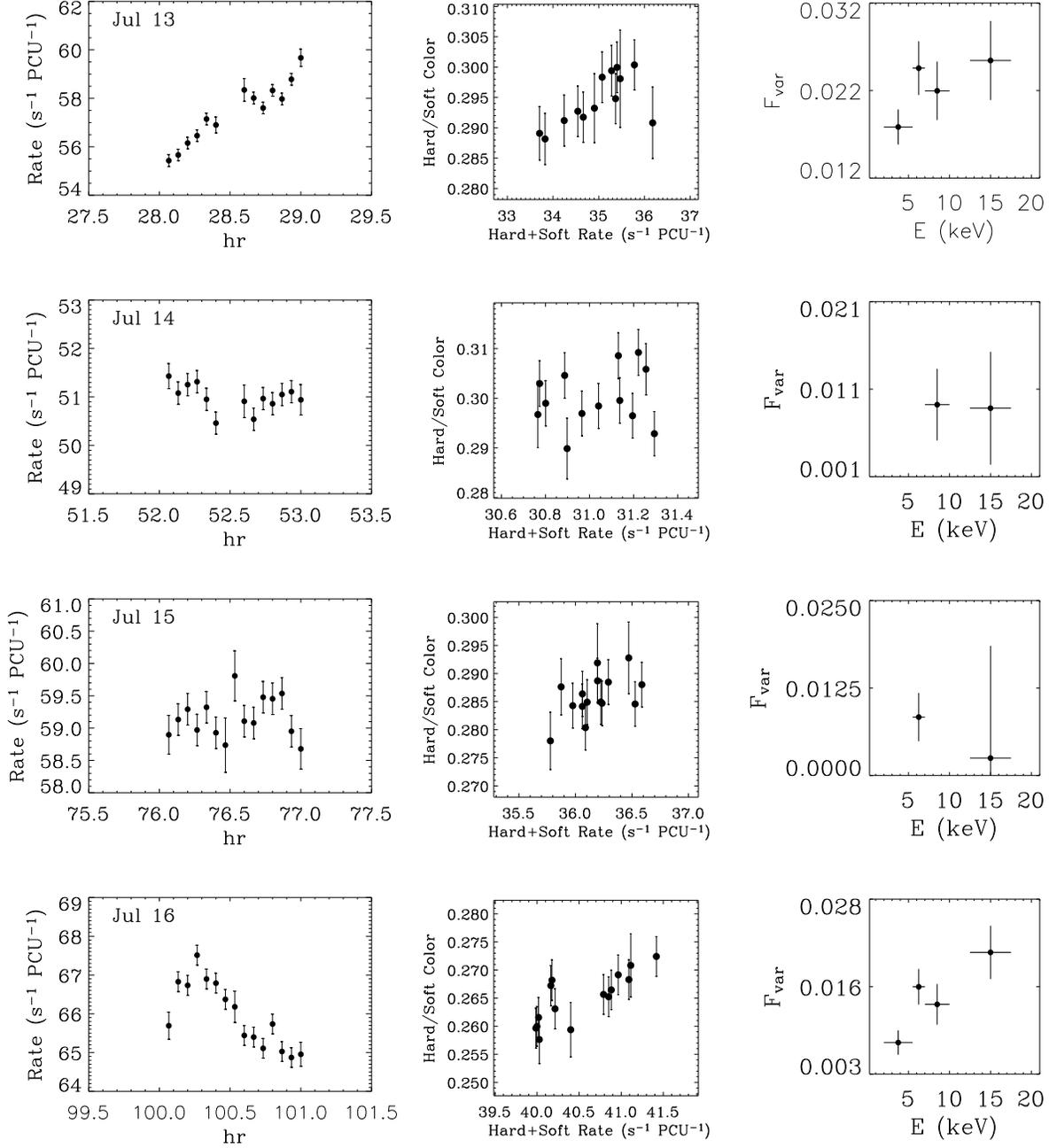}
\caption{{\it First column:} X-ray light curves of the longest pointings in July 1997.
Time bins are 240 s. {\it Second column:} X-ray color [10-20 keV]/[2-5 keV] plotted versus count rate for each individual pointing. {\it Third column:} Fractional variability parameter plotted versus the energy for each individual pointing in July 1997.} 
\label{figure:fig11}
\end{figure}


\begin{table} 
\caption{Time-resolved X-ray spectroscopy in July 1997}
\begin{center}
\begin{tabular}{lcccccc} 
\hline        
\hline
\noalign{\smallskip}        
Date & Exp.[s] & rate [c/s] & $\Gamma_1$ & $E_{\rm break}$ [keV]  & $\Gamma_2$ & norm \\
\noalign{\smallskip}  
\hline  
\noalign{\smallskip}
07/13/97 & 2688 & $252.3\pm0.3$ & $1.73_{-0.04}^{+0.03}$ & $6.98_{-0.74}^{+1.27}$ & $1.88_{-0.02}^{+0.03}$ & $0.15_{-0.01}^{+0.01}$\\
\noalign{\smallskip}
\hline
07/14/97 & 2848 & $224.7\pm0.3$ & $1.71_{-0.04}^{+0.03}$ & $6.75_{-0.76}^{+1.49}$ & $1.86_{-0.01}^{+0.01}$ & $0.13_{-0.01}^{+0.01}$\\
\noalign{\smallskip}
\hline  
07/15/97 & 3120 & $259.9\pm0.3$ & $1.75_{-0.05}^{+0.03}$ & $6.46_{-0.64}^{+0.88}$ & $1.90_{-0.01}^{+0.02}$ & $0.16_{-0.01}^{+0.01}$\\
\noalign{\smallskip}
\hline 
07/16/97 & 3264 & $289.1\pm0.3$ & $1.80_{-0.04}^{+0.03}$ & $6.78_{-0.58}^{+0.92}$ & $1.99_{-0.02}^{+0.02}$ & $0.20_{-0.01}^{+0.01}$\\
\noalign{\smallskip}
\hline
\end{tabular}
\end{center}
\label{tab3}
\end{table}       

\end{document}